\documentclass[sigconf]{acmart}

\usepackage{xspace,balance,tabularx,multirow}
\usepackage{flushend}
\usepackage{tikz}
\usepackage{pgfplots}
\pgfplotsset{compat=1.16}
\usepackage{subcaption}
\usepackage[ruled, vlined, linesnumbered]{algorithm2e}
\usepackage{xcolor}
\usepackage{colortbl}
\usepackage{bbold}
\SetKwComment{Comment}{$\triangleright$\ }{}
\usepackage{enumitem}

\AtBeginDocument{%
  \providecommand\BibTeX{{%
    \normalfont B\kern-0.5em{\scshape i\kern-0.25em b}\kern-0.8em\TeX}}}

\setcopyright{none}

\renewcommand\footnotetextcopyrightpermission[1]{} 
\newcommand{\argmax}[1]{\underset{#1}{\operatorname{arg}\,\operatorname{max}}\;}
\newcommand{\argmin}[1]{\underset{#1}{\operatorname{arg}\,\operatorname{min}}\;}

\makeatletter
\newcommand*\bigcdot{\mathpalette\bigcdot@{.5}}
\newcommand*\bigcdot@[2]{\mathbin{\vcenter{\hbox{\scalebox{#2}{$\m@th#1\bullet$}}}}}
\makeatother

\def\done{\hspace*{\fill} {$\square$}}
\def\header{\vspace{1mm} \noindent}

\newcommand{\ie}{{\it i.e.},\xspace}
\newcommand{\eg}{{\it e.g.},\xspace}

\newcommand\lnorm[1]{\left\lVert#1\right\rVert}

\newcommand{\AM}{\mathbf{A}\xspace}
\newcommand{\DM}{\mathbf{D}\xspace}
\newcommand{\XM}{\mathbf{X}\xspace}
\newcommand{\RM}{\mathbf{R}\xspace}
\newcommand{\IM}{\mathbf{I}\xspace}
\newcommand{\SM}{\mathbf{S}\xspace}
\newcommand{\FM}{\mathbf{F}\xspace}
\newcommand{\PtM}{\mathbf{P}_V\xspace}
\newcommand{\PaM}{\mathbf{P}_R\xspace}
\newcommand{\PM}{\mathbf{P}\xspace}

\newcommand{\HM}{\mathbf{H}\xspace}
\newcommand{\ZM}{\mathbf{Z}\xspace}
\newcommand{\YM}{\mathbf{Y}\xspace}
\newcommand{\MM}{\mathbf{M}\xspace}

\newcommand{\kmeans}{$k\textrm{-}\mathsf{Means}$\xspace}
\newcommand{\asc}{$\mathsf{USC}$\xspace}
\newcommand{\gasc}{$\mathsf{ACMin}$\xspace}
\newcommand{\gascb}{$\boldsymbol{\mathsf{ACMin}}$\xspace}

\newlength\lengtha 

\definecolor{forestgreen}{RGB}{34, 139, 34}

\definecolor{RYB1}{RGB}{192, 128, 255}
\definecolor{RYB2}{RGB}{255, 192, 32}
\definecolor{RYB3}{RGB}{139, 0, 0}
\definecolor{RYB4}{RGB}{0, 128, 255}

\newenvironment{customlegend}[1][]{%
    \begingroup
    \csname pgfplots@init@cleared@structures\endcsname
    \pgfplotsset{#1}%
}{%
    \csname pgfplots@createlegend\endcsname
    \endgroup
}%

\def\addlegendimage{\csname pgfplots@addlegendimage\endcsname}

\makeatletter
\newcommand\footnoteref[1]{\protected@xdef\@thefnmark{\ref{#1}}\@footnotemark}
\makeatother

\begin{document}

\title{Effective and Scalable Clustering on Massive Attributed Graphs}\titlenote{This is the full version of the paper appearing in TheWebConf 2021.}
\subtitle{Technical Report}

\author{Renchi Yang}
\affiliation{%
  \institution{Nanyang Technological University}
   \country{}
}
\email{rcyang@ntu.edu.sg}

\author{Jieming Shi}
\authornote{Corresponding author.}
\affiliation{%
  \institution{Hong Kong Polytechnic University}
  \country{}
}
\email{jieming.shi@polyu.edu.hk}

\author{Yin Yang}
\affiliation{%
  \institution{Hamad bin Khalifa University}
  \country{}
}
\email{yyang@hbku.edu.qa}

\author{Keke Huang}
\affiliation{%
  \institution{National University of Singapore}
  \country{}
}
\email{kkhuang@nus.edu.sg}

\author{Shiqi Zhang}
\affiliation{%
  \institution{National University of Singapore}
  \country{}
}
\email{s-zhang@comp.nus.edu.sg}

\author{Xiaokui Xiao}
\affiliation{%
  \institution{National University of Singapore}
  \country{}
}
\email{xkxiao@nus.edu.sg}

\renewcommand{\shortauthors}{Yang and Shi, et al.}

\begin{abstract}
Given a graph $G$ where each node is associated with a set of attributes, and a parameter $k$ specifying the number of output clusters, {\em $k$-attributed graph clustering} ($k$-AGC) groups nodes in $G$ into $k$ disjoint clusters, such that nodes within the same cluster share similar topological and attribute characteristics, while those in different clusters are dissimilar. This problem is challenging on massive graphs, \textit{e.g.}, with millions of nodes and billions of 
attribute values.
For such graphs, existing solutions either incur prohibitively high costs, or produce clustering results with compromised quality. 

In this paper, we propose \gasc, an efficient approach to $k$-AGC that yields high-quality clusters with costs linear to the size of the input graph $G$. The main contributions of \gasc are twofold: (i) a novel formulation of the $k$-AGC problem based on an \textit{attributed multi-hop conductance} quality measure custom-made for this problem setting, which effectively captures cluster coherence in terms of both topological proximities and attribute similarities, and (ii) a linear-time optimization solver that obtains high quality clusters iteratively, based on efficient matrix operations such as orthogonal iterations, an alternative optimization approach, as well as an initialization technique that significantly speeds up the convergence of \gasc in practice.

Extensive experiments, comparing 11 competitors on 6 real datasets, demonstrate that \gasc consistently outperforms all competitors in terms of result quality measured against ground truth labels, while being up to orders of magnitude faster. In particular, on the Microsoft Academic Knowledge Graph dataset with 265.2 million edges and 1.1 billion attribute values, \gasc outputs high-quality results for 5-AGC within 1.68 hours using a single CPU core, while none of the 11 competitors finish within 3 days.
\end{abstract}
\settopmatter{printfolios=true}
\maketitle

\section{Introduction}\label{sec:intro}
Node clustering is a fundamental task in graph mining \cite{schaeffer2007graph,lancichinetti2009community,ng2002spectral,yang2019efficient}, and finds important real-world applications, \eg community detection in social networks \cite{fortunato2010community}, functional cartography of metabolic networks \cite{guimera2005functional}, and protein grouping in biological networks \cite{voevodski2009finding}. Traditionally, node clustering is done based on the graph topology, \textit{i.e.}, by grouping together well-connected nodes. This approach, however, is often insufficient to obtain high-quality clusters \cite{freeman1996cliques,hric2014community}, especially when the graph comes with attributes associated to nodes. In such \textit{attributed graphs}, well-connected nodes tend to share similar attributes; meanwhile, nodes with similar attributes are also likely to be well-connected, as observed in \cite{la2010randomization,kossinets2006empirical}.  Therefore, to obtain high-quality node clustering, it is important to consider both graph topology and node attributes. The resulting \textit{attributed graph clustering} has use cases such as gene clustering in biological networks \cite{hanisch2002co}, group-oriented marketing in communication networks \cite{xu2012model}, service/app recommendation, and online advertising in social networks \cite{huang2020community,liji2018improved}.


This paper focuses on {$k$-\em attributed graph clustering} ($k$-AGC), which takes as input an attributed graph $G$ and a parameter $k$, and aims to partition $G$ into $k$ disjoint node clusters
$C_1,C_2,\cdots,C_k$, such that the nodes within the same cluster $C_i$ are not only well-connected to each other, but also share similar attribute values, whereas the nodes in different clusters are distant to each other and share less attributes. It is highly challenging to devise a $k$-AGC algorithm that yields high-quality clusters, especially on massive graphs, \eg with millions of nodes and billions of 
attribute values. Most existing solutions (\eg \cite{combe2012combining,meng2018coupled,neville2003clustering,steinhaeuser2008community,ruan2013efficient,zhou2009graph,zhou2010clustering,nawaz2015intra,falih2017anca,xu2012model,zhang2019attributed,wang2019attributed,li2018community,wang2017mgae,akbas2017attributed,yang2009combining}) fail to scale to such large graphs, since they either incur prohibitive computational overhead, or produce clustering results with compromised quality. For instance, a common methodology \cite{zhou2009graph,combe2012combining,nawaz2015intra,falih2017anca} relies on materializing the attribute similarity between every pair of nodes in the input graph $G$, and, thus, requires $O(n^2)$ space for $n$ nodes, which is infeasible for a graph with numerous nodes. Methods based on probabilistic models (\eg \cite{yang2009combining,zanghi2010clustering,nowicki2001estimation,xu2012model,he2017joint})
generally require immense costs on large graphs to estimate the likelihood parameters in their respective optimization programs. Among the faster solutions, some (\eg \cite{combe2012combining,meng2018coupled,neville2003clustering,steinhaeuser2008community,ruan2013efficient}) reduce the problem to non-attributed graph clustering by re-weighting each edge $(u,v)$ in $G$ based on the attribute similarity between nodes $u$ and $v$. This approach, however, ignores attribute similarities between nodes that are not directly connected, and, consequently, suffers from severe result quality degradation.
Finally, $k$-AGC could be done by first applying attributed network embedding to the input graph (\eg \cite{yang2015network,yang2018binarized,ijcai2019-low,liu2018content,meng2019co,zhou2018prre,hamilton2017inductive,yangscale2021}) to obtain an embedding vector for each node, and subsequently feeding the resulting embeddings to a non-graph method such as \kmeans clustering \cite{hartigan1979algorithm,park2009simple}. This two-stage pipeline leads to sub-optimal result quality, however, since the node embedding methods do not specifically target for graph clustering, as demonstrated in our experiments.

Facing the challenge of $k$-AGC on massive attributed graphs, we propose \gasc (short for \underline{A}ttributed multi-hop \underline{C}onductance \underline{Min}imization), a novel solution that seamlessly incorporates both graph topology and node attributes to identify high-quality clusters, while being highly scalable and efficient on massive graphs with numerous nodes, edges and attributes.
Specifically, \gasc computes $k$-AGC by solving an optimization problem, in which the main objective is formulated based on a novel concept called \textit{average attributed multi-hop conductance}, which is a non-trivial extension to conductance \cite{chung1997spectral,yang2019efficient}, a classic measure of node cluster coherence. The main idea is to map both node relationships (\ie connections via edges) and similarities (\ie common attributes) to motions of a random walker. Then, we show that the corresponding concept of conductance in our setting, \ie attributed multi-hop conductance, is equivalent to the probability that a random walker starting from a node in a cluster (say, $C$) terminates at any node outside the cluster $C$. Accordingly, our goal is to identify a node partitioning scheme that minimizes the average attributed multi-hop conductance among all $k$ clusters in the result.


Finding the exact solution to the above optimization problem turns out to be infeasible for large graphs, as we prove its NP-hardness. 
Hence, \gasc tackles the problem via an approximate solution with space and time costs linear to the size of the input graph. 
In particular, there are three key techniques in the \gasc algorithm.
First, instead of actually sampling random walks, \gasc
converts the optimization objective into its equivalent matrix form, and iteratively refines a solution via efficient matrix operations, \ie orthogonal iterations \cite{rutishauser1969computational}.
Second, the \gasc solver applies an alternative optimization approach and randomized SVD \cite{halko2011finding} to efficiently generate and refine clustering results. Third, \gasc includes an effective greedy initialization technique that significantly speeds up the convergence of the iterative process in practice. 

We formally analyze the asymptotic time and space complexities of \gasc, and evaluate its performance thoroughly by comparing against 11 existing solutions on 6 real datasets. The quality of a clustering method's outputs is evaluated by both (i) comparing them with ground truth labels, and (ii) measuring their attributed multi-hop conductance, which turns out to agree with (i) on all datasets in the experiments. The evaluation results demonstrate that \gasc consistently outperforms its competitors in terms of clustering quality, at a fraction of their costs. In particular, on the {\em Flickr} dataset, the performance gap between \gasc and 
the best competitor is as large as 28.6 percentage points, measured as accuracy with respect to ground truth. On the Microsoft Academic Knowledge Graph ({\em MAG}) dataset with 265.2 million edges and 1.1 billion attribute values, \gasc terminates in 1.68 hours for a 5-AGC task, while none of the 11 competitors finish within 3 days.

The rest of this paper is organized as follows. Section \ref{sec:so} presents our formulation of the $k$-AGC problem, based on two novel concepts: attributed random walks and attributed multi-hop conductance. Section \ref{sec:solutionoverview} overviews the proposed solution \gasc and provides the intuitions of the algorithm. Section \ref{sec:mainalgo} describes the complete \gasc algorithm and analyzes its asymptotic complexity. Section \ref{sec:exp} contains an extensive set of experimental evaluations. Section \ref{sec:relatedwork} reviews related work, and Section \ref{sec:ccl} concludes the paper with future directions.

\vspace{-1mm}
\section{Problem Formulation}\label{sec:so}


Section \ref{sec:preliminary} provides necessary background and defines common notations. Section \ref{sec:arwmodel} describes a random walk model that incorporates both topological proximity and attribute similarity information. Section \ref{sec:acmodel} defines the novel concept of attributed multi-hop conductance, which forms the basis of the objective function in our $k$-AGC problem formulation, presented in Section \ref{sec:objfunction}.

\vspace{-1mm}
\subsection{Preliminaries}\label{sec:preliminary}

\begin{table}[!t]
\centering
\renewcommand{\arraystretch}{1.1}
\begin{scriptsize}
\caption{Frequently used notations.}\vspace{-3mm} \label{tbl:notations}
\resizebox{\columnwidth}{!}{%
	\begin{tabular}{|p{0.7in}|p{2.15in}|}
		\hline
		{\bf Notation} &  {\bf Description}\\
		\hline
		$G$=$(V,E_{V},R,E_{R})$   & A graph $G$ with node set $V$, edge set $E_{V}$, attribute set $R$, and node-attribute association set $E_{R}$.\\
		\hline
		$n, d$   & The number of nodes (\ie $|V|$) and the number of attributes (\ie $|R|$) in $G$, respectively.\\
        \hline
		$k$   & The number of clusters. \\
		\hline
		$\AM, \DM, \RM$   & The adjacency, out-degree and attribute matrices of $G$. \\
		\hline
		$ \PtM, \PaM$ & The topological transition and attributed transition matrices of $G$, respectively.\\
		\hline
		$\alpha,\beta$ & Stopping and attributed branching probabilities. \\
		\hline
		$\SM$ & The attributed random walk probability matrix (see Eq. \eqref{eq:sexact}).\\
		\hline
		$\FM$ & The top-$k$ eigenvectors of $\SM$.\\
		\hline
		$\YM, \Psi(\YM)$ & A $k\times n$ node-cluster indicator (\ie NCI) and the average attributed multi-hop conductance (\ie AAMC) of $\YM$ (see Eq. \eqref{eq:objt}).\\
		\hline
	\end{tabular}%
}
\end{scriptsize}
\vspace{0mm}
\end{table}

Let $G=(V, E_V, R, E_R)$ be an {\it attributed graph} consisting of a node set $V$ with cardinality $n$, a set of edges $E_V$ of size $m$, each connecting two nodes in $V$, a set of attributes\footnote{Following common practice in the literature \cite{xu2012model,yangscale2021}, we assume that the attributes have already been pre-processed, \eg  categorical attributes such as marital status are one-hot encoded into binary ones.} $R$ with cardinality $d$, and a set of node-attribute associations $E_R$, where each element is a tuple $(v_i ,r_j, w_{i,j})$ signifying that node $v_i \in V$ is directly associated with attribute $r_j \in R$ with a weight $w_{i,j}$. Without loss of generality, we assume that each edge $(v_i, v_j) \in E_V$ is directed; an undirected edge $(v_i, v_j)$ is simply converted to a pair of directed edges with opposing directions $(v_i, v_j)$ and $(v_j, v_i)$.
A high-level definition of the $k$-AGC problem is as follows.

\begin{definition}[$k$-Attributed Graph Clustering \textnormal{($k$-AGC) \cite{zhou2009graph}}] \label{def:kagc}
Given an attributed graph $G$ and the number $k$ of clusters, $k$-AGC aims to partition the node set $V$ of $G$ into disjoint subsets: $C_1,C_2,\cdots,C_k$, such that (i) nodes within the same cluster $C_i$ are close to each other, while nodes between any two clusters $C_i,C_j$ are distant from each other; and (ii) nodes within the same cluster $C_i$ have homogeneous attribute values, while the nodes in different clusters may have diverse attribute values.
\end{definition}

Note that the above definition does not include a concrete optimization objective that quantifies node proximity and attribute homogeneity. As explained in Sections \ref{sec:arwmodel}-\ref{sec:objfunction}, the design of effective cluster quality measures is non-trivial, and is a main contribution of this paper. The problem formulation is completed later in Section \ref{sec:objfunction} with a novel objective function.

Regarding notations, we denote matrices in bold uppercase, \eg $\MM$. We use $\MM[i]$ to denote the $i$-th row vector of $\MM$, and $\MM[:,j]$ to denote the $j$-th column vector of $\MM$. In addition, we use $\MM[i,j]$ to denote the element at the $i$-th row and $j$-th column of $\MM$. Given an index set $\mathcal{I}$, we let $\MM[\mathcal{I}]$ (resp.\ $\MM[:,\mathcal{I}]$) be the matrix block of $\MM$ that contains the row (resp.\ column) vectors of the indices in $\mathcal{I}$.
 

Let $\AM$ be the adjacency matrix of the input graph $G$, \ie $\AM[v_i, v_j] = 1$ if  $(v_i,v_j)\in E_V$, otherwise $\AM[v_i, v_j] = 0$. Let $\DM$ be the diagonal out-degree matrix of $G$, \ie $\DM[v_i,v_i] = \sum_{v_j\in V}{\AM[v_i,v_j]}$. We define the topological transition matrix of $G$ as $\PtM = \DM^{-1}\AM$. 
Furthermore, we define an attribute matrix $\RM \in \mathbb{R}^{n\times d}$, such that $\RM[v_i,r_j] = w_{i,j}$ is the weight associated with the entry ($v_i$, $r_j$, $w_{ij}$) $ \in E_R$. We refer to $\RM[v_i]$ as node $v_i$'s \textit{attribute vector}. 
Also, let $d_{out}(v_i)$ and $d_{in}(v_i)$ represent the out-degree and in-degree of node $v_i$ in $G$, respectively.
Table \ref{tbl:notations} lists the frequently used notations throughout the paper.

\begin{figure}[!t]
\centering
\begin{small}
\subfloat[Clusters $C_1,C_2$ ]{\includegraphics[width=0.4\columnwidth]{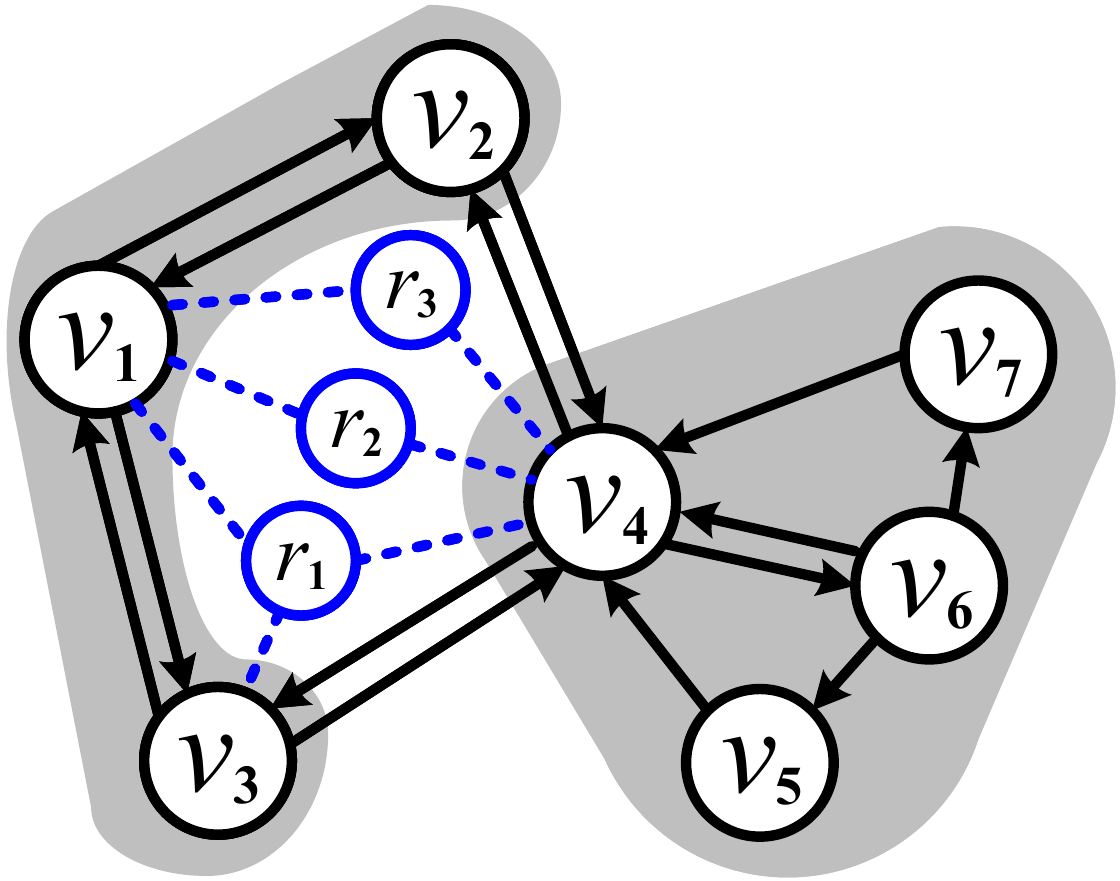}\label{fig:toya}}
\hspace{2mm}
\subfloat[Clusters $C_1^{\prime},C_2^{\prime}$]{\includegraphics[width=0.38\columnwidth]{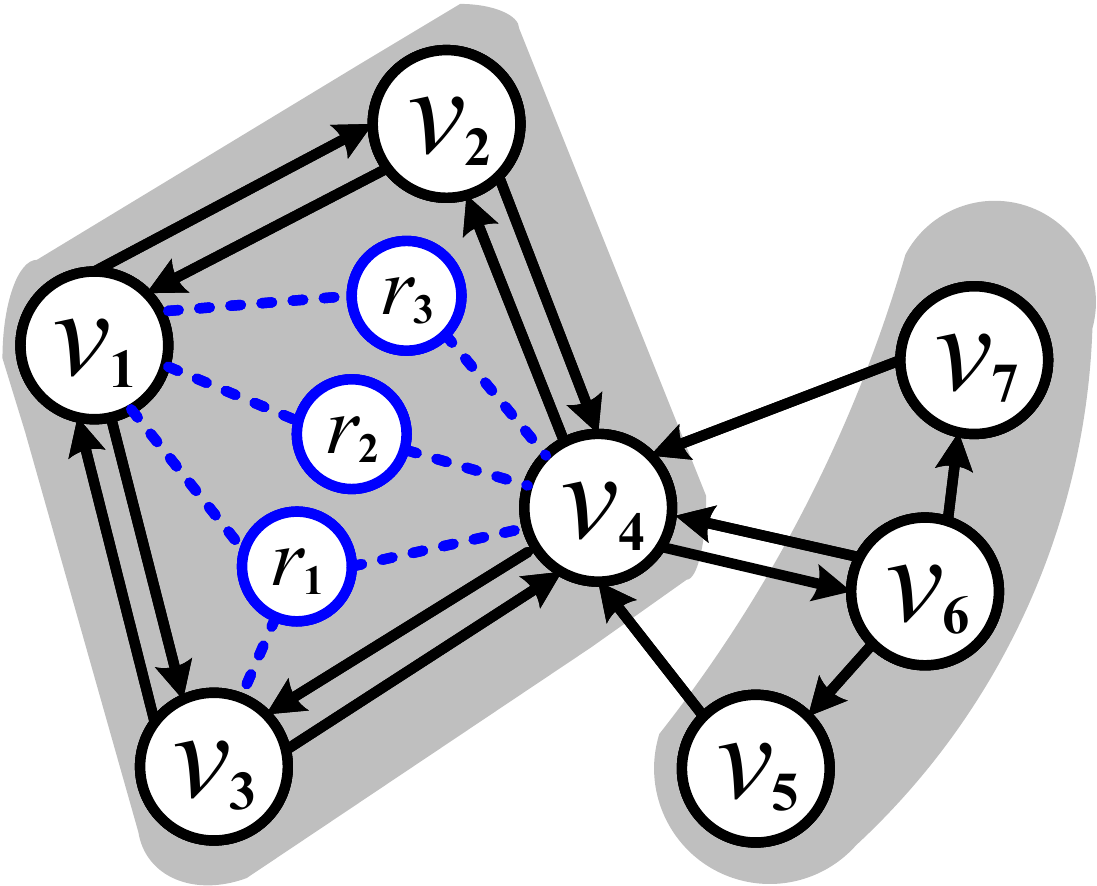}\label{fig:toyb}}
\vspace{-3mm}
\caption{Example attributed graph and clustering schemes.}\label{fig:toy}
\vspace{-1mm}
\end{small}
\vspace{-1mm}
\end{figure}

\vspace{-1mm}
\subsection{Attributed Random Walk Model} \label{sec:arwmodel}
Random walk is an effective model for capturing multi-hop relationships between nodes in a graph \cite{lovasz1993random}. Common definitions of random walk, \eg random walk with restart (RWR) \cite{jeh2003scaling,tong2006fast}, consider only graph topology but not node attributes. Hence, we devise a new \textit{attributed random walk} model that seamlessly integrates topological proximity and attribute similarity between nodes in a coherent framework, which plays a key role in our formulation of the $k$-AGC problem, elaborated later.



Given an attributed graph $G$, we first define the \textit{attributed transition probability} and \textit{topological transition probability} between a pair of nodes $v_i$ and $v_j$ in $G$. We say that $v_i$ and $v_j$ are connected via attribute $r_x$, iff. $v_i$ and $v_j$ have a common attribute $r_x$. For example, in Figure \ref{fig:toy}, nodes $v_1$ and $v_4$ are connected via three attributes $r_1-r_3$ (shown in blue dashed lines).
The attributed transition probability from $v_i$ to $v_j$ via $r_x$ is defined as $\frac{\RM[v_i,r_x]\cdot\RM[v_j,r_x]}{\sum_{v_l\in V}\sum_{r_y\in R}{\RM[v_i,r_y]\cdot{\RM[v_l,r_y]}}}$, which corresponds to the motion of the random walker that hops from $v_i$ to $v_j$ through a ``bridge'' $r_x$.
Accordingly, we define the attributed transition probability matrix $\PaM$ of $G$ as:

\begin{equation}\label{eq:pam}
\textstyle \PaM[v_i,v_j]=\frac{\RM[v_i]\cdot\RM[v_j]^{\top}}{\sum_{v_l\in V}{\RM[v_i]\cdot\RM[v_l]^{\top}}}.
\end{equation}

Intuitively, $\PaM[v_i,v_j]$ models the attributed transition probability from $v_i$ to $v_j$ via any attribute in $R$.

Meanwhile, following conventional random walk definitions, for any two nodes $v_i$ and $v_j$ that are directly connected by an edge in $G$, \ie $(v_i,v_j) \in E_V$, the topological transition probability $\PtM[v_i,v_j]$ from $v_i$ to $v_j$ is $\frac{1}{d_{out}(v_i)}$, where $d_{out}(v_i)$ is the out-degree of node $v_i$. The topological transition matrix $\PtM$ can then be obtained by $\PtM = \DM^{-1}\AM$, where $\DM$ and $\AM$ are the node degree and adjacency matrices of $G$, respectively. Based on the above concepts, we formally define attributed random walk as follows.

\begin{definition}[Attributed Random Walk]\label{def:arw}
Given an attributed graph $G$, a stopping probability $\alpha\in (0,1)$, and an attributed branching probability $\beta\in (0,1)$, an attributed random walk starting from node $v_i$ in $G$ performs one of the following actions at each step:

\begin{enumerate}[leftmargin=*]
\item with probability $\alpha$, stop at the current node (denoted as $v_j$),
\item with probability $1-\alpha$, jump to another node $v_l$ as follows:
\begin{enumerate}
\item (attributed transition) with probability $\beta$, jump to another node $v_l$ via any attribute with probability $\PaM[v_j,v_l]$,
\item (topological transition) with probability $1-\beta$, jump to an out-neighbor $v_l$ of $v_j$ with probability $\PtM[v_j,v_l]$.


\end{enumerate}
\end{enumerate}
\end{definition}


Based on Definition \ref{def:arw}, the following lemma\footnote{All proofs appear in Appendix \ref{sec:proofs}} shows how to directly compute the probability $\SM[v_i,v_j]$ that an attributed random walk starting from node $v_i$ stops at node $v_j$.

\begin{lemma}\label{lem:st}
Given an attributed graph $G$, the probability that an attributed random walk starting from node $v_i$ stops at node $v_j$ is
\begin{equation}\label{eq:sexact}
\textstyle\SM[v_i,v_j]=\alpha\sum_{\ell=0}^{\infty}{(1-\alpha)^{\ell}\cdot((1-\beta)\cdot\PtM+\beta\cdot\PaM)^\ell}[v_i,v_j].
\end{equation}
\end{lemma}

Note that computing $\SM$ directly using Eq. \eqref{eq:sexact} is inefficient, which involves sampling numerous attributed random walks. Instead, the proposed solution \gasc, presented later, computes the probabilities in $\SM$ based on an alternative matrix representation, without simulating any attributed random walk.

\vspace{-1mm}
\subsection{Attributed Multi-Hop Conductance} \label{sec:acmodel}
Conductance is widely used to evaluate the quality of a node cluster in a graph \cite{chung1997spectral,yang2019efficient}. A smaller conductance indicates a more coherent cluster, and vice versa. Specifically, given a cluster $C$ of graph $G$, the conductance of $C$, denoted as $\widehat{\Phi}(C)$, is defined as follows.
\begin{equation}\label{eq:classiccond}
\textstyle \widehat{\Phi}(C)=\frac{|\textrm{cut}(C)|}{\min\{\textrm{vol}(C), \textrm{vol}(V \backslash C)\}},
\end{equation}
where $\textrm{vol}(C)=\sum_{v_i\in C}{d_{out}(v_i)}$, \ie the sum of the out-degrees of all nodes in $C$, and $\textrm{cut}(C)=\{(v_i,v_j) \mid v_i\in C, v_j\in {V\backslash C}\}$, \ie the set of outgoing edges with an endpoint in $C$ and the other in ${V\backslash C}$. Intuitively, $\widehat{\Phi}(C)$ is smaller when $C$ has fewer outgoing edges linking to the nodes outside the cluster (\ie lower inter-cluster connectivity), and more edges with both endpoints within $C$ (higher intra-cluster connectivity).


In our setting, the classic definition of conductance $\widehat{\Phi}(C)$ is inadequate, since it captures neither attribute information nor multi-hop relationships between nodes. Figure \ref{fig:toy} illustrates an example in which $\widehat{\Phi}(C)$ leads to counter-intuitive cluster quality measurements. The example contains nodes $v_1$-$v_7$ and attributes $r_1$-$r_3$. 
Suppose that we aim to partition $G$ into two clusters. As shown in Figure \ref{fig:toyb}, node $v_4$ is mutually connected to nodes $v_2$ and $v_3$, and also shares many attributes (\ie $r_1$, $r_2$, and $r_3$) and neighbors (\ie $v_2$ and $v_3$) with node $v_1$; in contrast, among nodes $v_5$-$v_7$, $v_4$ is only mutually connected to $v_6$, and share no common attributes with them. Imagine that this is in a social media setting where each node represents a user, and each edge indicates a follow relationship; then, $v_4$ is clearly closer to nodes $v_1$-$v_3$ than to nodes $v_5$-$v_7$, due to its stronger connections and shared attributes to the former group. 
However, the conductance definition in Eq. \eqref{eq:classiccond} leads to the counter-intuitive conclusion that favors the clustering scheme $C_1=\{v_1,v_2,v_3\}$ and $C_2=\{v_4,v_5,v_6,v_7\}$ in Figure \ref{fig:toya} over $C_1'$ and $C_2'$ in Figure \ref{fig:toyb}, since the conductance $\widehat{\Phi}(C_1)=\widehat{\Phi}(C_2)=\frac{1}{3}\le \widehat{\Phi}(C_1^{\prime})=\widehat{\Phi}(C_2^{\prime})=\frac{2}{5}$.


To address the above issue, we propose a new measure of cluster quality dubbed {\em attributed multi-hop conductance}, which can be viewed as an adaptation of conductance to the problem setting of $k$-AGC. 
Specifically, given a cluster $C$ of an attributed graph $G$, suppose that we perform $n_r$ attributed random walks from each node $v_i$ in $C$. Let $w(v_i,v_j)$ be the number of walks from $v_i$ stopping at $v_j$. Then, we can use the following quantity instead of Eq. \eqref{eq:classiccond} as a measure of cluster coherence:
\begin{align*}
\textstyle \mathbb{E}\left[\frac{\sum_{v_i\in C, v_j\in V\backslash C}{w(v_i,r_j)}}{n_r\cdot |C|}\right]=\frac{\sum_{v_i\in C, v_j\in V\setminus C}{\mathbb{E}[\frac{w(v_i,v_j)}{n_r}}]}{|C|}.
\end{align*}
Intuitively, the above value quantifies the expected portion of the attributed random walks \textit{escaping} from $C$, \ie stopping at any outside node $v_j\in V\backslash C$. Hence, the smaller the number of escaped walks, the higher the cluster coherence. Further, observe that $\mathbb{E}[\frac{w(v_i,v_j)}{n_r}]$ corresponds to the probability that an attributed random walk starting from $v_i$ terminates at $v_j$, \ie $\SM[v_i,v_j]$ in Eq. \eqref{eq:sexact}.  Accordingly, we arrive at the following definition of attributed multi-hop conductance $\Phi(C)$. 
\begin{definition}[Attributed Multi-Hop Conductance]\label{def:amc}
Given a cluster $C$ of an attributed graph $G$, the attributed multi-hop conductance $\Phi(C)$ of the cluster $C$ is defined as
\begin{equation} \label{eq:attrConductance}
\textstyle\Phi(C)=\sum_{v_i\in C, v_j\in V\setminus C}{\frac{\SM[v_i,v_j]}{|C|}}.
\end{equation}
\end{definition}


\subsection{Objective Function}\label{sec:objfunction}

Given an input attributed graph $G$, we aim to partition all nodes into $k$ disjoint clusters $C_1,C_2,\cdots,C_k$, such that their {\em average attributed multi-hop conductance} (AAMC) $\phi$ of the $k$ clusters is minimized, as follows.
\begin{equation}\label{eq:objc}
\textstyle\phi^{*}=\min_{C_1,C_2,\cdots,C_k}\frac{{\sum_{i=1}^{k}{\Phi(C_i)}}}{k}.
\end{equation}

The above objective, in combination with Definition \ref{def:kagc}, completes our formuation of the $k$-AGC problem. As an example, in Figure \ref{fig:toy}, let $\alpha=0.2,\beta=0.5$. Then, we have $\Phi(C_1)=0.121,\Phi(C_2)=0.125$ for the clusters $C_1,C_2$ in Figure \ref{fig:toya}, and $\Phi(C_1^{\prime})=0.025, \Phi(C_2^{\prime})=0.185$ for the clusters $C_1^{\prime},C_2^{\prime}$ in Figure \ref{fig:toyb}. The AAMC values of these two clustering results are $\frac{\Phi(C_1)+\Phi(C_2)}{2}=0.123>\frac{\Phi(C_1^{\prime})+\Phi(C_2^{\prime})}{2}=0.105$, which indicate that $C_1^{\prime},C_2^{\prime}$ are a better clustering of $G$, which agrees with our intuition explained in Section \ref{sec:acmodel}.

\vspace{-1mm}
\section{Solution Overview} \label{sec:solutionoverview}
This section provides a high-level overview of the proposed solution \gasc for $k$-AGC computation, and explains the intuitions behind the algorithm design. The complete \gasc method is elaborated later in Section \ref{sec:mainalgo}.

First, we transform the optimization objective in Eq. \eqref{eq:objc} to an equivalent form that is easier to analyze. For this purpose, we introduce the following binary node-cluster indicator (NCI) $\YM\in \mathbb{1}^{k\times n}$ to represent a clustering result:
\begin{equation}\label{eq:y}
\textstyle \YM[C_i,v_j]=\begin{cases}
\textstyle 1 \quad&\textstyle \text{$v_j \in C_i$,}
\\
\textstyle 0 \quad&\textstyle \text{$v_j \in V \setminus C_i$},
\end{cases}
\end{equation}
where $C_i$ is the $i$-th cluster and $v_j$ is the $j$-th node in the node set $V$ of the input graph $G$.
Based on NCI $\YM$, the following lemma presents an equivalent form of the AAMC objective function in Eq. \eqref{eq:objc}.


\begin{lemma}\label{lem:y-c}
Given a clustering result $C_1,C_2,\cdots,C_k$, represented by NCI $\YM$, the AAMC of $C_1,C_2,\cdots,C_k$ can be obtained by:
\begin{equation} \label{eq:phi-y}
\textstyle\frac{\sum_{i=1}^{k}\Phi(C_i)}{k}=\frac{2}{k}\cdot\textnormal{trace}(((\YM\YM^{\top})^{-\frac{1}{2}} \YM)\cdot(\IM-\SM)\cdot((\YM\YM^{\top})^{-\frac{1}{2}} \YM)^{\top}) 
\end{equation}
\end{lemma}


Then, our optimization objective for $k$-AMC is transformed to: 
\begin{align}
&\quad\quad\quad\quad\quad\quad\quad\quad\quad\quad\phi^{*}=\min_{\YM\in \mathbb{1}^{k\times n}}{\Psi(\YM)} \label{eq:objt}\\
&\textrm{where } \Psi(\YM)=\textstyle\frac{2}{k}\cdot\textrm{trace}(((\YM\YM^{\top})^{-\frac{1}{2}} \YM)\cdot(\IM-\SM)\cdot((\YM\YM^{\top})^{-\frac{1}{2}} \YM)^{\top})\nonumber
\end{align}

Note that Eq. \eqref{eq:objt} is equivalent to Eq. \eqref{eq:objc}, and yet the former is more friendly to analysis. In particular, we have the following negative result.

\begin{lemma}\label{lem:np}
The optimization problem of finding the optimal values of $\YM$ from the objective function in Eq. \eqref{eq:objt} is NP-hard.\done
\end{lemma}


Accordingly, to devise a solution for $k$-AGC on massive graphs, we focus on approximate techniques for optimizing our objective. Observe that the NP-hardness of our objective function in Eq. \eqref{eq:objt} is due to the requirement that elements of the NCI are binary, \ie $\YM\in \mathbb{1}^{k\times n}$. Thus, we apply a common trick that relaxes NCI elements from binary to \textit{fractional}, \ie $\YM\in \mathbb{R}^{k\times n}$. The following lemma shows a sufficient condition to find the optimal values of fractional NCI $\YM$: when the row vectors of $(\YM\YM^{\top})^{-\frac{1}{2}}\YM$ are the top-$k$ eigenvectors of matrix $\SM$ (defined in Lemma \ref{lem:st}), \ie the $k$ eigenvectors corresponding to the $k$ largest eigenvalues of $\SM$. 

\begin{lemma}\label{lem:topk-eigen}
Assume that we relax the requirement $\YM\in \mathbb{1}^{k\times n}$ to $\YM\in \mathbb{R}^{k\times n}$. Let $\FM\in \mathbb{R}^{k\times n}$ denote the matrix consisting of the top-$k$ eigenvectors of $\SM$. Then, the optimal value of $\YM$ for the objective $\min_{\YM\in \mathbb{R}^{k\times n}}{\Psi(\YM)}$ is obtained when  $(\YM\YM^{\top})^{-\frac{1}{2}} \YM = \FM$, which leads to a value of $\Psi(\YM)$ no larger than the solution of the original optimal objective $\phi^*$ in Eq. \eqref{eq:objt}.\done
\end{lemma}
The optimal value of fractional $\YM$, however, does not directly correspond to a clustering solution, which requires the NCI to be binary. The following lemma points to a way to obtain a good approximation of the optimal binary $\YM\in \mathbb{1}^{k\times n}$.


\begin{lemma}\label{lem:gennci}
Given the top-$k$ eigenvectors $\FM$ of $\SM$, if we obtain a binary NCI $\YM$ that satisfies
\begin{align}\label{eq:obj_sr}
&\ \ \min{\|\XM\FM-(\YM\YM^{\top})^{-\frac{1}{2}} \YM\|^2_F}\quad \textrm{s.t. $\YM\in\mathbb{1}^{k\times n},\ \XM^{\top}\XM=\IM$},
\end{align}
then $\Psi(\YM)\rightarrow \phi^*$ in Eq. \eqref{eq:objt}.\done
\end{lemma}

Based on Lemmata \ref{lem:topk-eigen} and \ref{lem:gennci}, to approximate the optimal binary NCI $\YM$, we can first compute the top-$k$ eigenvectors $\FM$ of $\SM$, and then solve for the best NCI $\YM$ that optimizes Eq. \eqref{eq:obj_sr}. The proposed algorithm \gasc follows this two-step approach.

There remain two major challenges in realizing the above idea:

\begin{itemize}[leftmargin=*]
    \item How to compute $\FM$ for large graphs. Note that it is prohibitively expensive to compute $\FM$ directly by performing eigen-deomposition on a materialized matrix $\SM$ (defined in Eq. \eqref{eq:sexact}), which would consume $\Omega(n^2)$ space and $\Omega(n^2k)$ time.
    
    
    \item Given $\FM$, how to efficiently compute $\YM\in\mathbb{1}^{k\times n}$ based on Eq. \eqref{eq:obj_sr}, which in itself is a non-trivial optimization problem.
    
\end{itemize}


To address the above challenges, the proposed method \gasc contains three key techniques. First, to compute  the top-$k$ eigenvectors $\FM$ of $\SM$, \gasc employs a scalable, iterative process based on \textit{orthogonal iterations} \cite{rutishauser1969computational}, which does not need to materialize $\SM$. 
Second, to find the best NCI $\YM$, \gasc applies an \textit{alternative optimization approach} and \textit{randomized SVD} \cite{halko2011finding} to efficiently optimize Eq. \eqref{eq:obj_sr}. 
Third, to acclerate the above iterative processes, \gasc includes an effective greedy algorithm to compute a high-quality initial value of $\YM$, which significantly speeds up convergence in practice. Overall, \gasc only requires space and time linear to the size of the input graph $G$. The next section presents the detailed \gasc algorithm and complexity analysis.

\begin{algorithm}[!t]
\begin{small}
	\caption{\gasc}
	\label{alg:gasc}
	\KwIn{$G,k,\alpha,\beta$.}
	\KwOut{$\YM$.}
	Compute $\widehat{\RM}$ by Eq. \eqref{eq:rhat}\;
	$\YM_0 \gets \mathsf{InitNCI}(\PtM, \widehat{\RM}, \RM, \alpha, \beta)$\;
	$\textstyle \FM_0 \gets (\YM_0\YM_0^{\top})^{-\frac{1}{2}}\YM_0$\;
	$\YM \gets \YM_0$\;
	$\phi \gets \mathsf{AppoxAAMC}(\PtM, \widehat{\RM}, \RM, \alpha, \beta, \YM_0)$\;
	
	\For{$\ell\gets 1$ to $t_e$}{
	     $\ZM_{\ell} \gets (1-\beta)\cdot\PtM {\FM}^{\top}_{\ell-1}+\beta\cdot\widehat{\RM}(\RM^{\top} {\FM}^{\top}_{\ell-1})$\;
	     ${\FM}_{\ell}\gets \mathsf{QR}(\ZM_{\ell})$\;
	     \lIf{$\FM_{\ell}=\FM_{\ell-1}$}{
	         {\bf break}
	     }
    	$\YM_\ell \gets \mathsf{GenNCI}(\FM_{\ell})$\;
    	$\phi_\ell \gets \mathsf{AppoxAAMC}(\PtM, \widehat{\RM}, \RM, \alpha, \beta, \YM_\ell)$\;
         \lIf{$\phi_\ell < \phi$}{
             $\phi \gets \phi_\ell,\ \YM \gets \YM_\ell$
         }
	}
	\Return $\YM$\;
\end{small}
\end{algorithm}

\vspace{-1mm}
\section{Detailed \gascb Algorithm}\label{sec:mainalgo}

This section presents the detailed \gasc algorithm, shown in Algorithm \ref{alg:gasc}. In the following, Sections \ref{sec:topke}-\ref{sec:initnci} detail the three most important components of \gasc: the computation of top-$k$ eigenvectors $\FM$, binary NCI $\YM$, and a greedy initialization of $\YM$, respectively. Section \ref{sec:complete} summarizes the complete \gasc algorithm and analyzes its complexity.

\vspace{-1mm}
\subsection{Computing Top-$k$ Eigenvectors $\FM$}\label{sec:topke}
Recall from Section \ref{sec:solutionoverview} that \gasc follows a two-step strategy that first computes $\FM$, the top-$k$ eigenvectors of $\SM$ (Eq. \eqref{eq:sexact}). Since materializing $\SM$ is infeasible on large graphs, this subsection presents our iterative procedure for computing $\FM$ without materializing $\SM$, which corresponds to Lines 6-9 of Algorithm \ref{alg:gasc}.



First of all, the following lemma reduces the problem of computing $\FM$ to computing the top-$k$ eigenvectors of $(1-\beta)\cdot\PM_V+\beta \cdot \PM_R$.
\begin{lemma}\label{lem:topkeigen}
Let $\FM$ be the top-$k$ eigenvectors of $(1-\beta)\cdot \PtM+\beta\cdot \PaM$. Then, $\FM$ is also the top-$k$ eigenvectors of $\SM$.\done
\end{lemma}
Computing the exact top-$k$ eigenvectors of $(1-\beta)\cdot\PM_V+\beta \cdot \PM_R$ is still rather challenging, however, since materializing $\PM_R$ also requires $\Omega(n^2)$ space. To tackle this issue, \gasc applies \textit{orthogonal iterations} \cite{rutishauser1969computational}, as follows. First, \gasc computes a normalized attribute vector $\widehat{\RM}[v_i]$ for each node $v_i$ in the graph using the following equation, leading to matrix $\widehat{\RM}$ (Line 1 in Algorithm \ref{alg:gasc}).
\begin{equation}\label{eq:rhat}
\textstyle \widehat{\RM}[v_i]=  \frac{\RM[v_i]}{\RM[v_i]\cdot\mathbf{r}^{\top}} \ \forall{v_i\in V}, \textrm{where}\ \mathbf{r} = \sum_{v_j\in V}{\RM[v_j]}.
\end{equation}
Comparing above equation with Eq. \eqref{eq:pam}, it follows that $\PM_R = \widehat{\RM}\RM^{\top}$. Hence, $(1-\beta)\cdot\PM_V+\beta \cdot \PM_R$ in Lemma \ref{lem:topkeigen} can be transformed to $(1-\beta)\cdot\PM_V+\beta \cdot \widehat{\RM}\RM^{\top}$, eliminating the need to materialize $\PM_R$.

Next, suppose that we are currently at the start of the $\ell$-th iteration (Line 6 of Algorithm \ref{alg:gasc}) with $\FM_{\ell-1}$ obtained in previous iteration. Note that in the first iteration, $\FM_0$ is computed from an initial value $\YM_0$ of $\YM$, elaborated in Section \ref{sec:initnci}. \gasc computes $\ZM_\ell = ((1-\beta)\cdot\PM_V+\beta \cdot \widehat{\RM}\RM^{\top})\FM_{\ell-1}^{\top}=(1-\beta)\cdot\PM_V\FM_{\ell-1}^{\top}+\beta \cdot \widehat{\RM}\cdot(\RM^{\top}\FM_{\ell-1}^{\top})$ (Line 7 of the algorithm), which can be done in $O(k\cdot (|E_V|+|E_R|))$ time. Then, \gasc employs $\mathsf{QR}$ decomposition \cite{demmel1997applied} (Line 8) to decompose $\ZM_\ell$ into two matrices: $\FM_{\ell}$ and $\boldsymbol{\Lambda}_{\ell}$, such that  $\ZM_{\ell}=\FM^{\top}_{\ell}\cdot \boldsymbol{\Lambda}_{\ell}$, where $\boldsymbol{\Lambda}_{\ell}$ is an upper-triangular matrix, and $\FM_{\ell}$ is orthogonal (\ie $\FM_{\ell}\FM^{\top}_{\ell}=\IM$). Clearly, the $\mathsf{QR}$ decomposition step can be done in $O(nk^2)$ time, leading to $O(k\cdot (|E_V|+|E_R|)+nk^2)$ total time for one iteration in the computation of $\FM$.

Suppose that $\FM_\ell$ converges in iteration $\ell=t_c$, \ie $\FM_{t_c}$ is the same as $\FM_{t_c-1}$ (Line 9). Then, we have $\ZM_{\ell}=((1-\beta)\cdot\PM_V+\beta \cdot\PM_R)\FM_{t_c}^{\top}=\FM^{\top}_{t_c}\cdot \boldsymbol{\Lambda}_{t_c}$. Considering that $\boldsymbol{\Lambda}_{\ell}$ is an upper-triangular matrix, and $\FM_{\ell}$ is orthogonal (\ie $\FM_{\ell}\FM^{\top}_{\ell}=\IM$), according to \cite{rutishauser1969computational}, we conclude that $\FM_{t_c}$ is the top-$k$ eigenvectors of $(1-\beta)\cdot\PM_V+\beta \cdot \PM_R$ and the diagonal elements of $\boldsymbol{\Lambda}_{t_c}$ are the top-$k$ eigenvalues. According to Lemma \ref{lem:topkeigen}, the row vectors of $\FM_{t_c}$ are also the top-$k$ eigenvectors of $\SM$.

Note that throughout the process for computing $\FM$, there is no materialization of either $\SM$ or $\PM_R$, which avoids the corresponding quadratic space requirement. Meanwhile, with a constant $k$, each iteration takes time linear to the size of the input graph $G$, which is far more scalable than decomposing $\SM$ directly. In practice, the number of required iterations can be significantly reduced through a good initialization, detailed later in Section \ref{sec:initnci}.




\vspace{-1mm}
\subsection{Computing Binary NCI $\YM$}\label{sec:gennci}
\begin{algorithm}[!t]
\begin{small}
	\caption{$\mathsf{GenNCI}$}
	\label{alg:discrete}
	\KwIn{$\FM$.}
	\KwOut{${\YM}$.}
	$\XM^{\prime}\gets \IM, \XM \gets \IM$\;
    \For{$\ell\gets 1$ to $t_m$}{
       \lFor{$i\gets 1$ to $k$}{
           Compute $\gamma_i$ by Eq.~\eqref{eq:gamma-k}
       }
       \For{$v_j\in V$}{
           Pick $c_i$ by Eq.~\eqref{eq:update-y}\;
           $\YM[:,v_j]\gets \mathbf{0}, \ \YM[c_i,v_j]\gets 1$\;
       }
       $\mathbf{U}, \boldsymbol{\Sigma}, \mathbf{V}\gets \mathsf{SVD}((\YM\YM^{\top})^{-\frac{1}{2}}\YM\FM^{\top})$\;
       $\XM^{\prime}\gets \XM,\ \XM \gets \mathbf{U}\cdot \mathbf{V}^{\top}$\;
       \lIf{$\XM=\XM^{\prime}$}{{\bf break}}
    }
	\Return ${\YM}$\;
\end{small}
\end{algorithm}

As described in Section \ref{sec:solutionoverview}, after obtaining the top-$k$ eigenvectors $\FM$ of $\SM$, \gasc proceeds to compute the binary NCI $\YM$ by solving the optimization problem in Eq. \eqref{eq:obj_sr}. In Algorithm \ref{alg:gasc}, this is done in Lines 10-12. Note that in \gasc, the computation of $\YM$ is performed once in every iteration for computing $\FM$, rather than only once after the final value of $\FM$ is obtained. This is because our algorithm is approximate, and, thus, the final value of $\FM$ does not necessarily lead to the best clustering quality, measured by AAMC (Section \ref{sec:objfunction}). Hence, \gasc computes $\YM_\ell$ and the corresponding AAMC $\phi_\ell$ for each iteration $\ell$, and udpate the current best result $\YM$ and $\phi$ whenever a better result is found (Lines 11-12 in Algorithm \ref{alg:gasc}).


Next we clarify the $\mathsf{GenNCI}$ function, shown in in Algorithm \ref{alg:discrete}, which computes the binary NCI $\YM_\ell\in \mathbb{1}^{k\times n}$ with $\FM_\ell$ in the current iteration $\ell$.
First, based on properties of matrix trace, we transform the optimization objective in Eq. \eqref{eq:obj_sr}, as follows.
\begin{align}
\|\XM\FM-(\YM\YM^{\top})^{-\frac{1}{2}} \YM\|^2_F
& = 2k-2\cdot\textrm{trace}((\YM\YM^{\top})^{-\frac{1}{2}}\YM\FM^{\top}\XM^{\top}).\label{eq:obj_srt}
\end{align}
$\mathsf{GenNCI}$ applies an alternative optimization approach \cite{stella2003multiclass} to minimize Eq. \eqref{eq:obj_srt}. Specifically, the algorithm updates two variables, $\XM$ and $\YM$ in an alternating fashion, each time fixing one of them and updating the other, according to the following rules.


\header
{\bf Updating $\YM$ with $\XM$ fixed.} Given $\FM$, according to Eq.~\eqref{eq:obj_srt}, with $\XM$ fixed, the function to optimize becomes:
\begin{equation}\label{eq:obj_ry}
\max_{\YM\in \mathbb{1}^{K\times  n}}{\textrm{trace}((\YM\YM^{\top})^{-\frac{1}{2}}\YM\FM^{\top}\XM^{\top})}
\end{equation}
Let $\MM = \FM^{\top}\XM^{\top}$. Eq.~\eqref{eq:obj_ry} is equivalent to
\begin{equation}
\textstyle\max_{\YM\in \mathbb{1}^{k\times n}}{\sum_{v_j\in V}{\sum_{i=1}^{k}{\left(\YM[c_i,v_j]\cdot\frac{\MM[v_j,c_i]}{\sqrt{\sum_{v_l\in V}{\YM[c_i,v_l]}}}\right)}}}.
\end{equation}
Since $\YM\in \mathbb{1}^{k\times n}$, for each column $\YM[:,v_j]$ ($v_j\in V$), we update the entry at $c_i$ of $\YM[:,v_j]$ (\ie $\YM[c_i,v_j]$) to 1, and 0 everywhere else, where $c_i$ is picked greedily as follows:
\begin{align}
&\textstyle c_i = \argmax{1\le c_l\le k}{\left[\frac{(1-\YM[c_l,v_j])\cdot\MM[v_j,c_l]}{\sqrt{\gamma^2_l+1}}+\frac{\YM[c_l,v_j]\cdot\MM[v_j,c_l]}{\gamma_l}\right]}\label{eq:update-y},\\
&\textstyle \quad\quad\quad\quad\quad\ \textrm{where}\ \gamma_l=\sqrt{\sum_{v_z\in V}{\YM[c_l,v_z]}}\label{eq:gamma-k},
\end{align}
meaning that we always update each column $\YM[:,v_j]$ ($v_j\in V$) such that the objective function in Eq. \eqref{eq:obj_ry} is maximized. Since both $\MM = \FM^{\top}\XM^{\top}$ and $\gamma_l$ can be precomputed at the beginning of each iteration, which takes $O(nk^2)$ time, it takes $O(nk)$ time to update the whole $\YM$ in each iteration.

\header
{\bf Updating $\XM$ with $\YM$ fixed.} Given $\FM$, according to Eq.~\eqref{eq:obj_srt}, with $\YM$ fixed, the function to optimize becomes:
\begin{equation}\label{eq:obj_rx}
\max_{\XM^{\top}\XM=\IM}{\textrm{trace}((\YM\YM^{\top})^{-\frac{1}{2}}\YM\FM^{\top}\XM^{\top})}
\end{equation}

The following lemma shows that the optimal $\XM$ in Eq. \eqref{eq:obj_rx} can be obtained via singular value decomposition (SVD) of matrix $(\YM\YM^{\top})^{-\frac{1}{2}}\YM\FM^{\top}$.
\begin{lemma}\label{lem:svd}
The optimal solution to the objective function in Eq.~\eqref{eq:obj_rx} is $\XM=\mathbf{U}\mathbf{V}^{\top}$, where $\mathbf{U}$ and $\mathbf{V}$ are the left and right singular vectors of $(\YM\YM^{\top})^{-\frac{1}{2}}\YM\FM^{\top}$ respectively.\done
\end{lemma}
To compute SVD of $(\YM\YM^{\top})^{-\frac{1}{2}}\YM\FM^{\top}\in \mathbb{R}^{k\times k}$, $\mathsf{GenNCI}$ employs the randomized SVD algorithm \cite{halko2011finding}, which finishes in $O(k^3)$ time.


With the above update rules for $\XM$ and $\YM$ respectively, $\mathsf{GenNCI}$ (Algorithm \ref{alg:discrete}) iteratively updates $\XM$ and $\YM$ for a maximum of $t_m$ iterations (Lines 2-9). In our experiments, we found that setting $t_m$ to 50 usually leads to satisfactory performance. Note that the iterations may converge earlier than $t_m$ iterations (Line 9). Since updating $\YM$ and $\XM$ takes $O(nk^2)$ and $O(k^3)$ time respectively, $\mathsf{GenNCI}$ terminates within $O(t_m\cdot (nk^2+k^3))$ time.


\vspace{-2mm}
\subsection{Effective NCI Initialization}\label{sec:initnci}

Next we clarify the computation of the initial value $\YM_0$ of the NCI (Line 2 of Algorithm \ref{alg:gasc}). If we simply assign random values to elements of $\YM_0$, the iterative process in \gasc from Lines 6 to 12 would converge slowly. 
To address this issue, we propose an effective greedy initialization technique $\mathsf{InitNCI}$,
which usually leads to fast convergence of \gasc in practice, as demonstrated in our experiments in Section \ref{sec:exp-init}.


Given a cluster $C$, recall that its attributed multi-hop conductance $\Phi(C)$ (Eq. \eqref{eq:attrConductance}) is defined based on the intuition that $\Phi(C)$ is lower when an attributed random walk from any nodes in $C$ is more likely to stop at a node within $C$.
Further, we observe that in practice, a high-quality cluster $C$ tends to have high intra-cluster connectivity via certain center nodes within $C$, and such a center node usually has \textit{high in-degree} (\ie many in-neighbors). In other words, the nodes belonging to the same cluster tend to have many paths to the center node of the cluster, and consequently, a random walk with restart (RWR) \cite{tong2006fast,jeh2003scaling} within a cluster is more likely to stop at the center node \cite{tabrizi2013personalized}. 
Based on these intuitions, we propose to leverage  graph topology (\ie $V$ and $E_V$ of the input attributed graph $G$) as well as RWR to quickly identify $k$ possible cluster center nodes, $V_\tau=\{v_{{\tau}_1},v_{{\tau}_2},\cdots,v_{{\tau}_k}\}\subset V$, and greedily initialize NCI $\YM_0$ by grouping the nodes in $V$ to a  center node according to their topological relationships to the center node.

\begin{algorithm}[!t]
\begin{small}
	\caption{$\mathsf{InitNCI}$}
	\label{alg:init}
	\KwIn{$\PtM, \alpha, \beta$.}
	\KwOut{$\YM_0$.}
    $\YM_0 \gets \mathbf{0}$, $V_\tau\gets\emptyset$\;
    $V_{\tau}'\gets \{v_{{\tau}_1},v_{{\tau}_2},\cdots,v_{{\tau}_{5k}}\}$ where $v_{\tau_i}$ is the node in $V$ with $i$-th largest in-degree\;
	$\boldsymbol{\Pi}_0 \gets \IM[:,V_{\tau}'],\ t\gets \frac{1}{\alpha}$\;
	\lFor{$\ell \gets 1$ to $t$}{
	$\boldsymbol{\Pi}_{\ell}\gets (1-\alpha)\cdot\PtM\boldsymbol{\Pi}_{\ell-1}+\boldsymbol{\Pi}_0$\
	}
	$\boldsymbol{\Pi}_{t} \gets \alpha\cdot \boldsymbol{\Pi}_{t}$\;
    \lFor {$v_\tau\in V_\tau'$} {compute $\sum_{v_j\in V}\boldsymbol{\Pi}_t[v_j,v_{\tau}]$}
    Select the top-$k$ nodes $v_\tau\in V_\tau'$ with the largest $\sum_{v_j\in V}\boldsymbol{\Pi}_t[v_j,v_{\tau}]$ into $V_\tau$ as the $k$ center nodes\;
	\lFor{$v_j\in V$}{select $v_{\tau_i}\in V_\tau$ with the largest $\boldsymbol{\Pi}_t[v_j,v_{\tau_i}]$, and set $\YM_0[i,v_j]\gets 1$}
	\Return $\YM_0$\;
\end{small}
\end{algorithm}

Algorithm \ref{alg:init} presents the pseudo-code of $\mathsf{InitNCI}$. 
After initializing $\YM_0$ to a $k\times n$ zero matrix and $V_\tau$ to an empty set at Line 1, the method first selects from $V$ a candidate set $V^{\prime}_{\tau}$ of size $5k$ (Line 2), which consists of the top-$(5k)$ nodes with the largest in-degrees. The nodes in $V^{\prime}_{\tau}$ serve as the candidate nodes for the $k$ center nodes to be detected. 
Then we compute the $t$-hop RWR value $\boldsymbol{\Pi}_t[v_j,v_{{\tau}}]$ from every node $v_j\in V$ to every node $v_\tau\in V^{\prime}_\tau$ from Lines 3 to 5 according to the following equation \cite{yang2020homogeneous}. 
\begin{equation}\label{eq:rwr}
\textstyle \boldsymbol{\Pi}_t=\sum_{\ell=0}^{t}{\alpha(1-\alpha)^{\ell}\PtM^{\ell}}\cdot\IM[:,V_{\tau}']
\end{equation}
In particular, we set $t=\frac{1}{\alpha}$ at Line 3, which is the expected length of an RWR, and is usually sufficient for our purpose. If $\boldsymbol{\Pi}_t[v_j,v_\tau]$ is large,  it means that the random walks starting from $v_j$ are more likely to stop at $v_\tau$, which matches our aforementioned intuition of possible cluster center nodes.

Then, at Line 6, for each candidate center node $v_\tau\in V_\tau'$, we compute the sum of $\boldsymbol{\Pi}_t[v_j,v_{\tau}]$ from all nodes $v_j\in V$ to $v_\tau$.
If $v_\tau$ has larger $\sum_{v_j\in V}\boldsymbol{\Pi}_t[v_j,v_{\tau}]$, it indicates that the random walks starting from any nodes in $V$ are more likely to stop at $v_\tau$.
Therefore, at Line 7, we select the top-$k$ nodes $v_\tau\in V_\tau'$ with the largest $\sum_{v_j\in V}\boldsymbol{\Pi}_t[v_j,v_{\tau_i}]$ as the $k$ possible center nodes in $V_\tau$.
At Line 8, for each node $v_j\in V$, we select the center node $v_{\tau_i}\in V_\tau$ with the largest $\boldsymbol{\Pi}_t[v_j,v_{\tau_i}]$ and greedily group $v_j$ and $v_{\tau_i}$ into the same $i$-th cluster by setting $\YM_0[i,v_j]$ to 1, completing the computation of $\YM_0$.

Note that Line 2 in Algorithm \ref{alg:init} takes $O(n+k\log(n))$ time, and the computation of $\boldsymbol{\Pi}_t$ requires $O(\frac{k}{\alpha}\cdot|E_V|)$ time. Therefore, $\mathsf{InitNCI}$ runs in $O(\frac{k}{\alpha}\cdot|E_V|)$ time.

\vspace{-2mm}
\subsection{Complete \gascb Algorithm and Analysis}\label{sec:complete}

\begin{algorithm}[!t]
\begin{small}
	\caption{$\mathsf{AppoxAAMC}$}
	\label{alg:phi}
	\KwIn{$\PtM, \widehat{\RM}, \RM, \alpha, \beta, \YM$.}
	\KwOut{$\phi$.}
     $\HM_0\gets (\YM\YM^{\top})^{-\frac{1}{2}}\YM,\ t\gets \frac{1}{\alpha}$\;
     \For{$\ell=1$ to $t$}{
          $\HM_{\ell} \gets (1-\alpha)\cdot ((1-\beta)\cdot\PtM \HM_{\ell-1}^{\top}+\beta\cdot \widehat{\RM}(\RM^{ \top} \HM_{\ell-1}^{\top}))+\HM_0$\;
     }
     $\phi \gets \frac{2}{k}\cdot\sum_{i=1}^{k}{\HM_0[i]\cdot(\HM^{\top}_0[i]-\alpha\cdot \HM_t[:,i])}$\;
	\Return $\phi$\;
\end{small}
\end{algorithm}

Algorithm \ref{alg:gasc} summarizes the pseudo-code of \gasc, which takes as input an attributed graph $G$, the number of clusters $k$,
random walk stopping probability $\alpha$, and attributed branching probability $\beta$ (defined in Definition \ref{def:arw}).
Initially (Line 1), \gasc computes matrix $\widehat{\RM}$, explained in Section \ref{sec:topke}.
Then (Line 2), \gasc computes an initial value $\YM_0$ for $\YM$ via $\mathsf{InitNCI}$ (Algorithm \ref{alg:init}), and derives the corresponding value $\FM_0$ for $\FM$ according to Lemma \ref{lem:topk-eigen} in Line 3.



Next (Line 5), we invoke $\mathsf{AppoxAAMC}$ (Algorithm \ref{alg:phi}) that uses $\YM$ to compute $\phi$, the best AAMC obtained so far. Note that the exact AAMC $\phi=\Psi(\YM)$ in Eq. \eqref{eq:objt} is hard to evaluate since $\SM$ in Eq. \eqref{eq:sexact} is the sum of an infinite series. Instead, $\mathsf{ApproxAAMC}$ performs a finite number $t=\frac{1}{\alpha}$ of iterations in Eq. \eqref{eq:sexact} to obtain an approximate AAMC, since the expected length of an attributed random walk is $\frac{1}{\alpha}$. Specifically, given $\PtM,\widehat{\RM},\RM,\alpha,\beta$ and $\YM$ as inputs, $\mathsf{AppoxAAMC}$ first initializes $\HM_0$ as $(\YM\YM^{\top})^{-\frac{1}{2}}\YM$ and the number of iterations $t$ to $\frac{1}{\alpha}$ (Line 1 of Algorithm \ref{alg:phi}). Then, it computes the intermediate result $\HM_t$ by $t$ iterations in Lines 2-3. 
Lastly $\mathsf{AppoxAAMC}$ computes $\phi$ with $\HM_t$ and $\HM_0$ at Line 4. 
Algorithm \ref{alg:phi} takes $O(\frac{k}{\alpha}\cdot(|E_V|+|E_R|))$ time with the precomputed $\widehat{\RM}$. 

Utilizing algorithms $\mathsf{GenNCI}$ and $\mathsf{AppoxAAMC}$, \gasc obtains the binary NCI $\YM_\ell$ and its corresponding quality measure $\phi_\ell$ for each iteration $\ell$, after obtaining $\FM_\ell$. 
\gasc may terminate upon convergence,
or reaching a preset maximum number of iterations $t_e$. In our experiments, we found that $t_e=200$ is usually sufficiently large for convergence.

Next we analyze the total time and space complexities of \gasc. 
The computation of $\widehat{\RM}$ at Line 1 in Algorithm \ref{alg:gasc} takes $O(|E_R|)$ time. Algorithm \ref{alg:phi} requires $O(nk+\frac{k}{\alpha}\cdot(|E_V|+|E_R|))$ time. 
In each iteration (Lines 6-12), Line 7 takes $O(k\cdot(|E_V|+|E_R|))$ time and the $\mathsf{QR}$ decomposition over $\ZM_{\ell}$ takes $O(nk^2)$ time. According to Section \ref{sec:gennci} and Section \ref{sec:initnci}, $\mathsf{GenNCI}$ and $\mathsf{InitNCI}$ run in $O(t_m\cdot (nk^2+k^3))$ and $O(\frac{k}{\alpha}\cdot|E_V|)$ time, respectively. Thus, the total time complexity of \gasc is $\textstyle O\left(k(\frac{1}{\alpha}+t_e)\cdot(|E_V|+|E_R|)+nk^2t_et_m+\frac{kt_e}{\alpha}\cdot|E_V|)\right)$ when $k \ll n$, which equals $O(|E_V|+|E_R|)$ when $t_e,t_m$ and $k$ are regarded as constants. The space overhead incurred by \gasc is determined by the storage of $\PtM,\widehat{\RM},\RM,\ZM_{\ell},\FM_{\ell}$ and $\HM_{\ell}$, which is bounded by $O(|E_V|+|E_R|+nk)$. 

\vspace{-1mm}
\section{Experiments}\label{sec:exp}
We experimentally evaluate \gasc against 11 competitors in terms of both clustering quality and efficiency on 6 real-world datasets. All experiments are conducted on a Linux machine powered by an Intel Xeon(R) Gold 6240@2.60GHz CPU and 377GB RAM. Source codes of all competitors are obtained from the respective authors.
\vspace{-1mm}
\subsection{Experimental Setup}
\begin{table}[t]
\centering
\renewcommand{\arraystretch}{1.2}
\begin{footnotesize}
\caption{Datasets. {\small (K=$10^3$, M=$10^6$, B=$10^9$)}}\label{tbl:exp-data}\vspace{-3mm}
\resizebox{\columnwidth}{!}{%
\label{fig:time-mag}\hspace{2mm}%
}%
\vspace{-4mm}
\end{small}
\caption{Running time with varying $k$ (best viewed in color).} \label{fig:time-all}
\vspace{-2mm}
\end{figure*}

\begin{table*}[!t]
\centering
\renewcommand{\arraystretch}{1.1}
\caption{CA, NMI and AAMC with ground-truth (Large CA, NMI, and small AAMC indicate high clustering quality).}\vspace{-3mm}
\begin{small}
\resizebox{\textwidth}{!}{%
\begin{tabular}{c|c c c|c c c|c c c| c c c| c c c|c c c}
\hline
\multirow{2}{*}{\bf Solution} & \multicolumn{3}{c|}{\bf {\em Cora}} & \multicolumn{3}{c|}{\bf {\em Citeseer}} & \multicolumn{3}{c|}{\bf {\em Pubmed}} & \multicolumn{3}{c|}{\bf {\em Flickr}} & \multicolumn{3}{c|}{\bf {\em TWeibo}} & \multicolumn{3}{c}{\bf {\em MAG-Scholar-C}}\\ \cline{2-19}
& CA & NMI & AAMC & CA & NMI & AAMC & CA & NMI & AAMC & CA & NMI & AAMC & CA & NMI & AAMC & CA & NMI & AAMC\\ 
\hline
Ground-truth  & 1.0 & 1.0 & 0.546 & 1.0 & 1.0 & 0.531 & 1.0 & 1.0 & 0.505 & 1.0 & 1.0 & 0.691 & 1.0 & 1.0 & 0.719 & 1.0 & 1.0 & 0.63 \\ 
\hline
$\mathsf{TADW}$  & 0.554 & 0.402 & 0.593 & 0.539 & 0.333 & 0.569 & 0.483 & 0.096 & 0.55 & 0.16 & 0.062 & 0.733 & - & - &- &   - & - & -\\
$\mathsf{LQANR}$  & 0.64 & 0.492 & 0.559 & 0.587 & 0.374 & 0.549 & 0.403 & 0.022 & 0.612 & 0.127  & 0.002  & 0.739 & - & - & -& -  &  - & -\\
$\mathsf{PRRE}$  & 0.547 & 0.396 & 0.604 & 0.576 & 0.322 & 0.592 & 0.62 & 0.269 & 0.518 & 0.454 &  0.321 & 0.713 & - & - &- & - & - & -\\
$\mathsf{PANE}$  & 0.601 & 0.462 & 0.577 & \underline{0.677} & \underline{0.421} & 0.537 & 0.618 & 0.252 & 0.512 & 0.402 & 0.265 & 0.708 & 0.215 & 0.004 & 0.752 & - & - & -\\
\hline
$\mathsf{CSM}$  & 0.308 & 0.149 & 0.612 & 0.247 & 0.11 & 0.615 & 0.393 & 0.022 & 0.565 & - & - &- & - & - &- & - & - & -\\
\hline
\end{tabular}
}
\end{small}
\label{tbl:acc-nmi}
\vspace{-3mm}
\end{table*}

\header
{\bf Datasets.} Table \ref{tbl:exp-data} shows the statistics of the 6 real-world directed attributed  graphs used in our experiments. $|V|$ and $|E_V|$ denote the number of nodes and edges, while $|R|$ and $|E_R|$ represent the number of attributes and node-attribute associations, respectively. $|C|$ is the number of ground-truth clusters in $G$. In particular, {\em Cora}\footnote{\label{fn:linqs}{\url{http://linqs.soe.ucsc.edu/data} (accessed October, 2020)}}, {\em Citeseer}\footnoteref{fn:linqs}, {\em Pubmed}\footnoteref{fn:linqs} and {\em MAG-Scholar-C}\footnote{\url{https://figshare.com/articles/dataset/mag_scholar/12696653} (accessed October, 2020)} are citation graphs, in which each node represents a paper and each edge denotes a citation relationship. {\em Flickr}\footnote{\url{https://github.com/xhuang31/LANE} (accessed October, 2020)} and {\em TWeibo} \footnote{\url{https://www.kaggle.com/c/kddcup2012-track1} (accessed October, 2020)} are social networks, in which each node represents a user, and each directed edge represents a following relationship.
Further, notice that all 6 datasets have ground-truth cluster labels, and the number of ground-truth clusters $|C|$ is also included in Table \ref{tbl:exp-data}.

\header
{\bf Competitors.} We compare \gasc with 11 competitors, including 7 $k$-AGC algorithms ($\mathsf{CSM}$ \cite{nawaz2015intra}, $\mathsf{SA\textrm{-}Cluster}$ \cite{zhou2009graph}, $\mathsf{BAGC}$ \cite{xu2012model}, $\mathsf{MGAE}$ \cite{wang2017mgae}, $\mathsf{CDE}$ \cite{li2018community}, $\mathsf{AGCC}$ \cite{zhang2019attributed}, \asc \cite{von2007tutorial}), and 4 recent attributed network embedding algorithms ($\mathsf{TADW}$ \cite{yang2015network}, $\mathsf{PANE}$ \cite{yangscale2021}, $\mathsf{LQANR}$ \cite{ijcai2019-low}, $\mathsf{PRRE}$ \cite{zhou2018prre}). The network embedding competitors are used together with \kmeans to produce clustering results. In addition, we also compare with the classic unnormalized spectral clustering method \asc \cite{von2007tutorial}, which directly works on $\SM$ to extract clusters by materializing $\SM$, computing the top-$k$ eigenvectors of $\SM$, and then applying \kmeans on the  eigenvectors.

\header
{\bf Parameter settings.} We adopt the default parameter settings of all competitors as suggested in their corresponding papers. Specifically, for attributed network embedding competitors, we set the embedding dimensionality to $128$. 
For \gasc, we set $t_e=200, t_m=50, \alpha=0.2$, and $\beta=0.35$. Competitor \asc shares the same parameter settings of $\alpha$, $\beta$, and $t_e$ with \gasc.

\header
{\bf Evaluation criteria.} For efficiency evaluation, we vary the number of clusters $k$ in $\{5,10,20,50,100\}$, and report the running time (seconds) of each method on each dataset in Section \ref{sec:exp-efficiency}. The reported running time does not include the time for loading datasets. We terminate a method if it fails to return results within 3 days. In terms of clustering quality, we report the proposed {\em AAMC} measure (\ie average attributed multi-hop conductance), {\em modularity} \cite{newman2004finding}, {\em CA} (clustering accuracy with respect to ground truth labels) and {\em NMI} (normalized mutual information) \cite{aggarwal2014data} to measure the clustering quality in Section \ref{sec:exp-quality}. 
Note that AAMC considers both graph topology and node attributes to measure clustering quality, while modularity only considers graph topology.
Also, note that CA and NMI rely on ground-truth clusters, while AAMC and modularity do not. Therefore, when evaluating by CA and NMI, we set $k$ to be $|C|$ as in Table \ref{tbl:exp-data} for each dataset; when evaluating by modularity and AAMC, we vary $k$ in $\{5,10,20,50,100\}$. 

\vspace{-2mm}
\subsection{Efficiency Evaluation}\label{sec:exp-efficiency}

\begin{figure*}[!t]
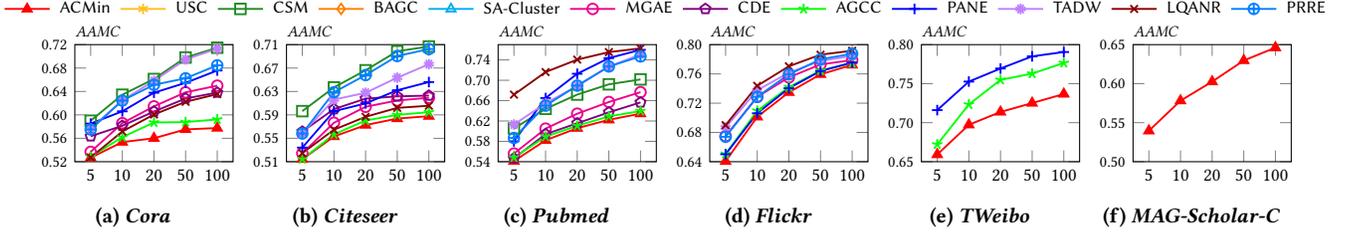

\centering
\begin{small}
\hspace{0.5mm}\label{fig:ac-mag}%
}%
\vspace{-3mm}
\end{small}
\caption{AAMC with varying $k$ (best viewed in color).} \label{fig:ac-all}
\vspace{-2mm}
\end{figure*}

\begin{figure*}[!t]
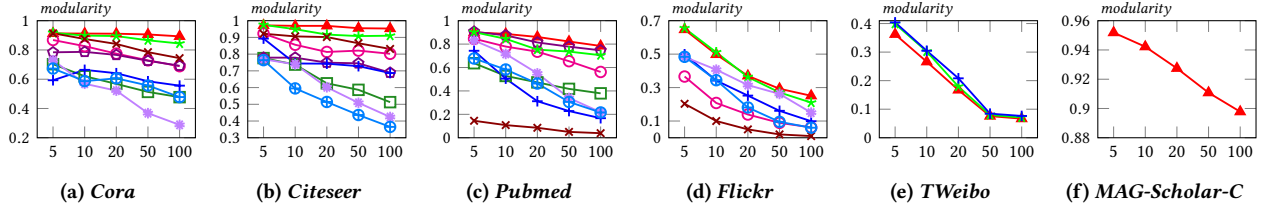

\centering
\begin{small}
\hspace{1mm}%
}%
\vspace{-3mm}
\end{small}
\caption{Modularity with varying $k$ (best viewed in color).} \label{fig:mod-all}
\vspace{-2mm}
\end{figure*}

\begin{figure*}[!t]
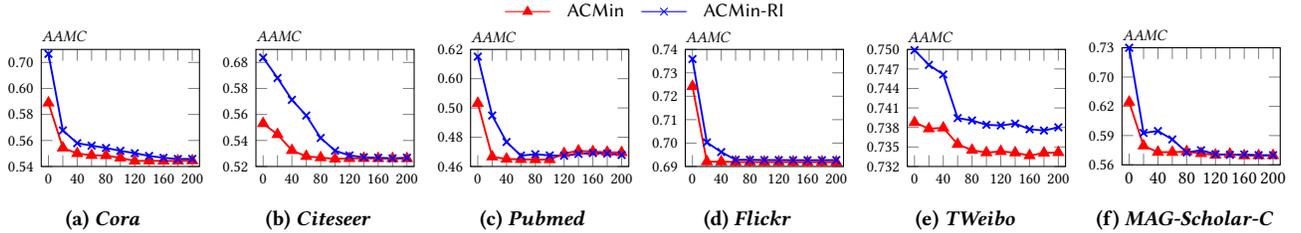

\centering
\begin{small}
%
\hspace{0mm}\label{fig:conv-mag}%
}%
\vspace{-3mm}
\end{small}
\caption{AAMC with varying $t_e$ (best viewed in color).} \label{fig:init-eff}
\vspace{-2mm}
\end{figure*}

Figure \ref{fig:time-all} presents the running time of all methods on all datasets when varying the number of clusters $k$ in $\{5,10,20,50,100\}$. The $y$-axis is the running time (seconds) in log-scale. As shown in Figure \ref{fig:time-all}, \gasc is consistently faster than all competitors on all datasets, often by up to orders of magnitude.
\gasc is highly efficient on large attributed graphs, \eg {\em TWeibo} and {\em MAG-Scholar-C} in Figures \ref{fig:time-twb} and \ref{fig:time-mag}, while most of the 11 competitors fail to return results within three days.
For instance, in Figure \ref{fig:time-twb}, when $k=5$, \gasc needs 630 seconds to finish, which is $7.4\times$ faster than $\mathsf{AGCC}$ (4634 seconds) and $71\times$ faster than $\mathsf{PANE}$ (44658 seconds), respectively.
Further, \gasc is the only method able to finish on {\em MAG-Scholar-C} dataset that has 265.2 million edges and 1.1 billion attribute values. Specifically, \gasc only needs 1.68 hours when $k=5$.
The high efficiency of \gasc on massive real datasets is due to the its high scalable algorithmic components, whose total cost is linear to the size of the input graph as analyzed in Section \ref{sec:complete}.
On small/moderate-sized attributed graphs in Figures \ref{fig:time-cora}-\ref{fig:time-flk}, \gasc is also significantly faster than the competitors, especially when $k$ is small. For instance, when $k=10$, on {\em Flickr} in Figure \ref{fig:time-flk}, \gasc takes 4 seconds, while the fastest competitor $\mathsf{PANE}$ needs 381 seconds. 
Note that the total running time of \gasc increases linearly with $k$, which is consistent with our time complexity analysis in Section \ref{sec:complete} when the number of edges $|E_V|+|E_R|$ far exceeds the number of nodes $n$.
The running time results for the 4 attributed network embedding competitors (\ie $\mathsf{TADW},\mathsf{LQANR},\mathsf{PRRE}$, and $\mathsf{PANE}$) are not sensitive to $k$, since their cost is dominated by the node embedding computation rather than \kmeans. Even on settings with a large $k$, \gasc is still faster than all these competitors.

\vspace{-2mm}
\subsection{Quality Evaluation}\label{sec:exp-quality}
\header
{\bf CA, NMI and AAMC with ground-truth.} Table \ref{tbl:acc-nmi} reports the CA, NMI, and AAMC scores of all methods, comparing to the ground-truth clusters of each dataset in Table \ref{tbl:exp-data}. We also report the AAMC values of the ground-truth labels, which are the lower than the AAMC obtained by all methods except for \gasc, which explicitly aims to minimize AAMC. Meanwhile, we observe that relative performance of all methods measured by AAMC generally agrees with CA. These results demonstrate that AAMC effectively measures effectively reflects clustering quality.

\gasc clearly and consistently achieves the best CA, NMI, and AAMC on all datasets. Specifically, on small attributed graphs, \ie {\em Cora}, {\em Citeseer} and {\em Pubmed}, compared with the best competitors (underlined in Table \ref{tbl:acc-nmi}) \gasc improves CA by $1.4\%$, $0.3\%$, and $2.3\%$, and NMI by $0.2\%$, $0.1\%$, and $3.6\%$, respectively. The CA and NMI of \gasc are also significantly better than the competitors on moderate-sized/large attributed graphs (\ie \textit{Flickr}, \textit{TWeibo}, and \textit{MAG-Scholar-C}). For instance, on {\em Flickr}, \gasc has CA $75.7\%$, which is $28.6\%$ higher than that of the best competitor $\mathsf{AGCC}$, which is only $47.1\%$.  
On {\em TWeibo}, \gasc is slightly better than $\mathsf{AGCC}$; note that on this dataset, \gasc is orders of magnitude faster than $\mathsf{AGCC}$ as shown in Figure \ref{fig:time-twb}. Hence, \gasc is overall preferable than $\mathsf{AGCC}$ in these settings. Finally, \gasc is the only $k$-AGC method capable of handling {\em MAG-Scholar-C}, and achieves CA $65.9\%$ and NMI $49.7\%$.
The superior clustering quality achieved by \gasc demonstrates the effectiveness of the proposed AAMC optimization objective in Section \ref{sec:so}, as well as our approximate solution for this optimization program described in Section \ref{sec:mainalgo}. 


\header
{\bf AAMC with varying $k$.} Figure \ref{fig:ac-all} reports the AAMC achieved by \gasc against all competitors on all datasets when varying the number $k$ of clusters in $\{5,10,20,50,100\}$. Observe that \gasc consistently produces the smallest AAMC under all $k$ settings on all datasets (smaller AAMC indicates better results), which confirms that the proposed \gasc algorithm (Algorithm \ref{alg:gasc}) effectively minimizes the proposed AAMC objective function defined in Section \ref{sec:objfunction}. 
In particular, as shown in Figure \ref{fig:ac-all}, when $k=10$, \gasc has AAMC better than the best competitor by a margin of $0.84\%$, $0.36\%$, $0.71\%$, $0.84\%$ and $2.6\%$ on {\em Cora}, {\em Citeseer}, {\em Pubmed}, {\em Flickr} and {\em TWeibo} respectively.
Figure \ref{fig:ac-mag} reports the AAMC achieved by \gasc on MAG-Scholar-C, which is the only method able to return results.
Further, considering the relative performance of all methods measured by CA and NMI generally agree with that measured by AAMC as shown in the results in Table \ref{tbl:acc-nmi}, and the fact that \gasc is far more efficient and scalable compared to its competitors as shown in Section \ref{sec:exp-efficiency},
we conclude that \gasc is the method of choice for $k$-AGC on massive graphs in practice.


\header
{\bf Modularity with varying $k$.} Figure \ref{fig:mod-all} reports the {\em modularity} of all methods on all datasets when varying  $k$ in $\{5,10,20,50,100\}$. Again, observe that, for all settings of $k$ and all datasets (except \textit{TWeibo}), \gasc has the highest modularity. 
In particular, \gasc obtains a substantial imporvement of up to $5\%, 4.7\%, 3.8\%$, and $4.1\%$ on {\em Cora}, {\em Citeseer}, {\em Pubmed} and {\em Flickr}, compared to the best competitor, respectively. 
Note that modularity only considers graph topology and ignores node attributes, indicating that modularity may not be able to fully evaluate clustering quality of \textit{attributed} graphs.
This may explain why on \textit{TWeibo} the modularity of \gasc is slightly lower than some competitors. Even so, \gasc still achieves high modularity under most cases, meaning that the proposed attributed random walk model can still preserve graph topological features for clustering, in addition to node attributes. 
\vspace{-1mm}
\subsection{Convergence Analysis of \gascb}\label{sec:exp-init}
In this section, we evaluate the convergence properties of \gasc, focusing on the effects of the greedy initialization technique $\mathsf{InitNCI}$ described in Section \ref{sec:initnci} on convergence speed. In particular, we compare \gasc with an ablated version $\mathsf{ACMin\textrm{-}RI}$ that replaces $\mathsf{InitNCI}$ at Line 2 of Algorithm \ref{alg:gasc} with random initialization of $\YM_0$.
The number $k$ of clusters to be detected is set to be $|C|$ as in Table \ref{tbl:exp-data} for each dataset.
Figure \ref{fig:init-eff} reports the AAMC (\ie $\Psi(\YM)$ in Eq. \eqref{eq:objt}) produced by \gasc and $\mathsf{ACMin\textrm{-}RI}$ per  iterations (Lines 6-12 in Algorithm \ref{alg:gasc}), when $t_e$ is set to $200$.
Observe that the AAMC produced by \gasc decreases significantly faster than that of $\mathsf{ACMin\textrm{-}RI}$ in the early iterations, and also converges faster than  $\mathsf{ACMin\textrm{-}RI}$. 
For instance, in Figure \ref{fig:conv-cite}, on \textit{Citeseer}, \gasc requires about 80 iterations to reach a plateaued AAMC, while $\mathsf{ACMin\textrm{-}RI}$ needs 140 iterations. Moreover, $\mathsf{GenNCI}$ is able to help \gasc to achieve lower AAMC at convergence as shown in Figure \ref{fig:init-eff}.
This experimental evaluation demonstrates the efficiency and effectiveness of the proposed greedy initialization technique in Section \ref{sec:initnci}.

\vspace{-3mm}
\section{Related Work}\label{sec:relatedwork}
Attributed graph clustering has been extensively studied in literature, as surveyed in \cite{bothorel2015clustering,falih2018community,chunaev2019community}. In the following, we review the existing methods that are most relevant to this work.

\header
{\bf Edge-weight-based clustering.} A classic methodology is to convert the input attributed graph to a weighted graph by assigning each edge a weight based on the attribute and topological similarity between the two nodes of the edge; then, traditional weighted graph clustering algorithms are directly applied \cite{combe2012combining,meng2018coupled,neville2003clustering,steinhaeuser2008community,ruan2013efficient}. For instance, Neville {\it et al.} \cite{neville2003clustering} assign a weight to each edge $(u,v)$ of the input attributed graph $G$ based on the number of attribute values that $u$ and $v$ have in common, and construct a weighted graph $G'$. Then they apply the classic spectral clustering \cite{von2007tutorial} over $G^{\prime}$ to produce clusters. However, these methods only consider the attributes of two directly connected nodes and use hand-crafted weights to represent attributes, and thus, result in inferior clustering quality. 

\header
{\bf Distance-based clustering.} Existing distance-based clustering solutions construct a distance matrix $\MM$ by combining the topological and attribute similarity between nodes, and then apply classic distance-based clustering methods, such as \kmeans \cite{hartigan1979algorithm} and $k\mathsf{\textrm{-}Medoids}$ \cite{park2009simple}, on $\MM$ to generate clusters.
For instance, $\mathsf{SA\textrm{-}Cluster}$ \cite{zhou2009graph} extends the original input attributed graph $G$ to an attribute-augmented graph $G'$ by treating each attribute as a node, and then samples random walks over $G^{\prime}$ to compute the distance between nodes in $G^{\prime}$, in order to construct $\MM$, which is then fed into a $k$-Centroids method to generate clusters. Further, $\mathsf{DCom}$ \cite{combe2012combining} applies hierarchical agglomerative clustering on a constructed distance matrix. $\mathsf{CSM}$ \cite{nawaz2015intra} computes the distance matrix $\MM$ based on a shortest path strategy that considers both structural and attribute relevance among nodes, and  applies $k\mathsf{\textrm{-}Medoids}$ over $\MM$ to generate clusters. $\mathsf{ANCA}$ \cite{falih2017anca} applies \kmeans for the sum of eigenvectors of the distance and similarity matrices to generate clusters. Distance-based clustering methods suffer from severe efficiency issues since they require to compute the distance of every node pair, resulting in $O(n^2)$ time and space overhead, which is prohibitive in practice. For instance, as shown in our experiments, both $\mathsf{SA\textrm{-}Cluster}$ and $\mathsf{CSM}$ suffer from  costly running time and poor clustering quality.

\header
{\bf Probabilistic-model-based clustering.} Based on the assumption that the  structure, attributes, and clusters of attributed graphs are generated according to a certain parametric distribution, there exist a collection of probabilistic-model-based clustering methods, which statistically infer a probabilistic model for attributed graph clustering, in order to generate clustering results. In particular, $\mathsf{PCL\textrm{-}DC}$ \cite{yang2009combining} combines a conditional model of node popularity and a discriminative model that reduces the impact of irrelevant attributes into a unified model, and then finds the clustering result that optimizes the model. $\mathsf{CohsMix}$ \cite{zanghi2010clustering} formulates the clustering problem by $\mathsf{MixNet}$ model \cite{nowicki2001estimation} and then utilizes a varient of EM algorithm to optimize it, in order to generate clustering results. $\mathsf{BAGC}$ \cite{xu2012model} designs a generative Bayesian model \cite{congdon2007bayesian} that produces a sample of all the possible combinations of a graph based on adjacency matrix $\AM$ and attribute matrix $\XM$, and aims to find a clustering result $C$ maximizing a conjoint probability $\mathbb{P}(C|\AM,\XM)$. 
Note that the optimization process to estimate the likelihood parameters in these probabilistic-model-based clustering methods often incurs substantial time overheads, as validated in our experiments (Section \ref{sec:exp-efficiency}).

\header
{\bf Embedding-based methods.} In recent years, 
a plethora of network embedding techniques
are proposed for attributed graphs. The objective of network embedding is to learn an embedding vector for each node such that the graph topology and attribute information surrounding the nodes can be preserved. We can directly employ traditional clustering methods (\eg \kmeans) over the embedding vectors to generate clusters \cite{hartigan1979algorithm,park2009simple}. 
$\mathsf{AA}\textrm{-}\mathsf{Cluster}$ \cite{akbas2017attributed} builds a weighted graph based on graph topology and node attributes, and then applies network embedding on the weighted graph to generate embeddings. $\mathsf{MGAE}$ \cite{wang2017mgae}  proposes a marginalized graph convolutional network to learn embeddings. $\mathsf{CDE}$ \cite{li2018community} learns node embeddings by optimizing a non-negative matrix factorization problem based on community structure embeddings and node attributes.
$\mathsf{DAEGC}$ \cite{wang2019attributed} fuses graph topology and node attributes via an attention-based autoencoder \cite{velivckovic2017graph} to obtain embeddings, and then generates soft labels to guide a self-training graph clustering procedure. 
$\mathsf{AGCC}$\cite{zhang2019attributed} utilizes 
an adaptive graph convolution method to learn embeddings, and then applies the spectral clustering on the similarity matrix computed from the learnt embeddings to obtain clusters. 
The above methods either incur immense overheads in learning embeddings or suffer from unsatisfactory clustering quality.
There are many attributed network embedding methods proposed, \eg \cite{yang2015network,yang2018binarized,ijcai2019-low,liu2018content,meng2019co,zhou2018prre,hamilton2017inductive,yangscale2021}. However, most of them are not specially designed for clustering purpose, leading to suboptimal clustering quality, as demonstrated in our experiments when comparing with $\mathsf{TADW}$, $\mathsf{LQANR}$, $\mathsf{PRRE}$ and $\mathsf{PANE}$. 

\vspace{-2mm}
\section{Conclusions}\label{sec:ccl}
This paper presents \gasc, an effective and scalable solution for $k$-AGC computation. \gasc achieves high scalability and effectiveness through a novel problem formulation based on the proposed attributed multi-hop conductance measure for cluster quality, as well as a carefully designed iterative optimization framework and an effective greedy clustering initialization method. Extensive experiments demonstrate that \gasc achieves substantial performance gains over the previous state of the art in terms of both efficiency and clustering quality. Regarding future work, we plan to study parallelized versions of \gasc, running on multi-core CPUs and GPUs, as well as in a distributed setting with multiple servers, in order to handle even larger datasets. Meanwhile, we intend to extend \gasc to handle attributed heterogeneous graphs with different types of nodes and edges.

\begin{acks}
This work is supported by the National University of Singapore under SUG grant R-252-000-686-133, by NPRP grant \#NPRP10-0208-170408 from the Qatar National Research Fund (a member of Qatar Foundation).
Jieming Shi is supported by the financial support(1-BE3T) of research project (P0033898) from Hong Kong Polytechnic University. The findings herein reflect the work, and are solely the responsibility, of the authors.
\end{acks}

\appendix
\section{Proofs}\label{sec:proofs}
\header
{\bf Proof of Lemma \ref{lem:st}.}
Let $p_{\ell}(v_i,v_j)$ be the probability that an attributed random walk starting from $v_i$ stops at $v_j$ at the $\ell$-th hop. We first prove that
\begin{equation}\label{eq:p-ell}
\textstyle p_{\ell}(v_i,v_j)=\alpha(1-\alpha)^{\ell}\cdot((1-\beta)\cdot\PtM+\beta\cdot\PaM)^\ell[v_i,v_j].
\end{equation}
Note that if Eq. \eqref{eq:p-ell} holds, the overall probability that an attributed random walk from $v_i$ terminates at $v_j$ is $\sum_{\ell=0}^{\infty}{p_\ell(v_i,v_j)}=\SM[v_i,v_j]$, which establishes the equivalence in Eq. \eqref{eq:sexact}. To this end, we  prove Eq. \eqref{eq:p-ell} by induction. First, let us consider the initial case that the attributed random walk terminates at source node $v_i$ with probability $\alpha$. In this case, $p_0(v_i,v_j)=\alpha$ if $v_i=v_j$; otherwise $p_0(v_i,v_j)=0$, which is identical to the r.h.s of Eq. \eqref{eq:p-ell} when $\ell=0$. Therefore, Eq. \eqref{eq:p-ell} holds when $\ell=0$. Assume that Eq. \eqref{eq:p-ell} holds at the $\ell^{\prime}$-th hop. Then the probability that an attributed random walk from $v_i$ visits any node $v_l\in V$ at the $\ell^{\prime}$-th hop is $(1-\alpha)^{\ell^{\prime}}\cdot((1-\beta)\cdot\PtM+\beta\cdot\PaM)^{\ell^{\prime}}[v_i,v_l]$. Based on this assumption, for the case $\ell=\ell^{\prime}+1$, with probability $1-\alpha$, it will navigate to node $v_j$ according to the probability $(1-\beta)\cdot\PtM[v_l,v_j]+\beta\cdot\PaM[v_l,v_j]$,
and finally stop at $v_j$ with probability $\alpha$. Thus, $\textstyle p_{\ell^{\prime}+1}(v_i,v_j) =\sum_{v_l\in V}{(1-\alpha)^{\ell^{\prime}}\cdot((1-\beta)\cdot\PtM+\beta\cdot\PaM)^{\ell^{\prime}}[v_i,v_l]} \cdot (1-\alpha)\alpha((1-\beta)\cdot\PtM[v_l,v_j]+\beta\cdot\PaM[v_l,v_j])=\textstyle \alpha(1-\alpha)^{\ell^{\prime}+1}\cdot((1-\beta)\cdot\PtM+\beta\cdot\PaM)^{\ell^{\prime}+1}[v_i,v_j]$, which completes the proof.\done

\header
{\bf Proof of Lemma \ref{lem:y-c}.}
By Eq. \eqref{eq:y}, for cluster $C_i$, we have vector $((\YM\YM^{\top})^{-\frac{1}{2}} \YM)[c_i]$, where each entry $((\YM\YM^{\top})^{-\frac{1}{2}} \YM)[c_i,v_j]={1}/{\sqrt{|C_i|}}$ if $v_j\in C_i$ and otherwise $((\YM\YM^{\top})^{-\frac{1}{2}} \YM)[c_i,v_j]=0$. Note that
\begin{align*}
&\textstyle 2\cdot((\YM\YM^{\top})^{-\frac{1}{2}} \YM)[c_i]\cdot(\IM - \SM)\cdot((\YM\YM^{\top})^{-\frac{1}{2}} \YM)[c_i]^{\top}\\
&\textstyle =\sum_{v_j,v_l\in V}{\SM[v_j,v_l]\cdot (((\YM\YM^{\top})^{-\frac{1}{2}} \YM)[c_i,v_j]-((\YM\YM^{\top})^{-\frac{1}{2}} \YM)[c_i,v_l])^2}\\
&\textstyle = \sum_{v_j\in C_i,v_l\in V\setminus C_i}{\SM[v_j,v_l]\cdot ((\YM\YM^{\top})^{-\frac{1}{2}} \YM)[c_i,v_j]^2} =\frac{\Phi(C_i)}{2}.
\end{align*}
Then we have
\begin{align*}
\textstyle\frac{\sum_{i=1}^{k}\Phi(C_i)}{k}&=\textstyle\frac{2}{k}\sum_{c_i=1}^{k}{((\YM\YM^{\top})^{-\frac{1}{2}} \YM)[c_i]\cdot(\IM - \SM)\cdot((\YM\YM^{\top})^{-\frac{1}{2}} \YM)[c_i]^{\top}}\\
&=\textstyle\frac{2}{k}\cdot\textrm{trace}(((\YM\YM^{\top})^{-\frac{1}{2}} \YM)\cdot(\IM-\SM)\cdot((\YM\YM^{\top})^{-\frac{1}{2}} \YM)^{\top}),
\end{align*}
which completes our proof.\done

\header
{\bf Proof of Lemma \ref{lem:np}.}
First, we construct a weighted graph $\mathcal{G}=(\mathcal{V},\mathcal{E})$ based on the input graph $G=(V,E_V,R,E_R)$ by letting $\mathcal{V}=V$ and $\mathcal{E}=\{(v_i,v_j,\SM[v_i,v_j]) \mid  v_i,v_j\in V\ \textrm{and}\ \SM[v_i,v_j]>0\}$, where $\SM[v_i,v_j]$ signifies the weight of edge $(v_i,v_j)$. Thus, Eq. \eqref{eq:objt} can be reduced to the objective function of the min-cut problem on $\mathcal{G}$, which is proven to be NP-hard in \cite{wagner1993betweenmin,goldschmidt1988polynomial}.\done

\header
{\bf Proof of Lemma \ref{lem:topk-eigen}.}
Let $\lambda_i(\MM)$ be the $i$-th smallest eigenvalue of matrix $\MM$. Note that $\ \forall{\YM}\in \mathbb{R}^{k\times n}$, $\textstyle ((\YM\YM^{\top})^{-\frac{1}{2}}\YM)\cdot ((\YM\YM^{\top})^{-\frac{1}{2}}\YM)^{\top}=\IM$, meaning that $\textstyle f(\YM)=((\YM\YM^{\top})^{-\frac{1}{2}}\YM)^{\top}\cdot ((\YM\YM^{\top})^{-\frac{1}{2}}\YM)$ is a projection matrix of rank $k$. Therefore, $\forall{i<n-k+1}, \lambda_i(f(\YM))=0$ and $\forall{i\ge n-k+1}, \lambda_i(f(\YM))=1$. By Von Neumann's trace inequality \cite{mirsky1975trace} and the property of matrix trace, for any $\YM\in \mathbb{R}^{k\times n}$, we have the following inequality:
\begin{align}
&\textstyle \Psi(\YM)= \frac{2}{k}\cdot\textrm{trace}(((\YM\YM^{\top})^{-\frac{1}{2}}\YM)\cdot(\IM-\SM)\cdot((\YM\YM^{\top})^{-\frac{1}{2}}\YM)^{\top})\nonumber\\
&\textstyle =\frac{2}{k}\cdot\textrm{trace}((\IM-\SM)\cdot f(\YM))\ge \frac{2}{k}\cdot\sum_{i=1}^{n}{\lambda_i(\IM-\SM)\cdot \lambda_{n-i+1}(f(\YM))}\nonumber\\
&\textstyle =\frac{2}{k}\cdot\sum_{i=1}^{k}\lambda_i(\IM-\SM)=\frac{2}{k}\cdot\sum_{i=1}^{k}(1-\lambda_{n-i+1}(\SM))\label{eq:psi-f-lb}.
\end{align}
Note that $\FM$ be the top-$k$ eigenvectors of $\SM$, implying $\FM\FM^{\top}=\IM$ and $(\IM-\SM)\cdot \FM[c_i]^{\top}=(1-\lambda_{n-i+1}(\SM))\cdot \FM[c_i]^{\top}$ for $1 \le i \le k$. Hence,
\begin{equation}\label{eq:frace}
\textstyle\frac{2}{k}\cdot\textrm{trace}(\FM(\IM-\SM)\FM^{\top})=\frac{2}{k}\cdot\sum_{i=1}^{k}{(1-\lambda_{n-i+1}(\SM))},
\end{equation}
which implies that $\Psi(\YM)$ is minimized when $((\YM\YM^{\top})^{-\frac{1}{2}}\YM)=\FM$.
Suppose $\YM^*\in \mathbb{1}^{k\times n}$ is the optimal solution to Eq. \eqref{eq:objt}. Therefore, with Eq. \eqref{eq:psi-f-lb} and Eq. \eqref{eq:frace}, the following inequality holds
\begin{align*}
\phi^*=\Psi(\YM^*)&=\frac{2}{k}\cdot\textrm{trace}(((\YM^*\YM^{*\top})^{-\frac{1}{2}} \YM^*)(\IM-\SM)((\YM^*\YM^{*\top})^{-\frac{1}{2}} \YM^*)^{\top})\\
& \textstyle \ge \frac{2}{k}\cdot\sum_{i=1}^{k}(1-\lambda_{n-i+1}(\SM))=\frac{2}{k}\cdot\textrm{trace}(\FM(\IM-\SM)\FM^{\top}),
\end{align*}
which finishes our proof.\done

\header
{\bf Proof of Lemma \ref{lem:gennci}.}
Eq. \eqref{eq:obj_sr} implies that $\XM\FM\rightarrow (\YM\YM^{\top})^{-\frac{1}{2}} \YM$ where $\XM^{\top}\XM=\IM$. By the property of matrix trace, we have
\begin{align}
\textstyle \Psi(\YM)&\textstyle =\frac{2}{k}\cdot\textrm{trace}((\YM\YM^{\top})^{-\frac{1}{2}} \YM\cdot(\IM-\SM)\cdot((\YM\YM^{\top})^{-\frac{1}{2}} \YM)^{\top})\nonumber\\ 
&\textstyle\rightarrow \frac{2}{k}\cdot\textrm{trace}(\XM\FM(\IM-\SM)\FM^{\top}\XM^{\top})\nonumber\\
&=\textstyle\frac{2}{k}\cdot\textrm{trace}(\XM^{\top}\XM\FM(\IM-\SM)\FM^{\top})=\frac{2}{k}\cdot\textrm{trace}(\FM(\IM-\SM)\FM^{\top})\label{eq:phiy}.
\end{align}
By Lemma \ref{lem:topk-eigen}, we have $\Psi(\YM) \rightarrow \phi^*$, completing our proof.\done

\header
{\bf Proof of Lemma \ref{lem:topkeigen}.}
We need the following lemmas for the proof.
\begin{lemma}[\cite{zhang2018arbitrary}]\label{thrm:lambda}
If $[\lambda,\mathbf{x}]$ is an eigen-pair of matrix $\mathbf{M}\in \mathbb{R}^{n\times n}$, then $[\sum_{\ell=0}^{t}{w_\ell\lambda^\ell},\mathbf{x}]$ is an eigen-pair of matrix $\sum_{\ell=0}^{t}{w_\ell\mathbf{M}^{\ell}}$.
\end{lemma}
\begin{lemma}[\cite{haveliwala2003second}]\label{lem:stoch-eigen}
Given a $\mathbf{M}\in \mathbb{R}^{n\times n}$ satisfying $\sum_{j=1}^{n}{\mathbf{M}[i,j]}=1\ \forall{1\le i\le n}$ and each entry $\mathbf{M}[i,j]\ge 0\ \forall{1\le i,j\le n}$, the largest eigenvalue $\lambda_1$ of $\mathbf{M}$ is 1.
\end{lemma}

Suppose that $[\lambda_i,\mathbf{x}_i]$ is an eigen-pair of $(1-\beta)\cdot\PtM+\beta\cdot\PaM$ and $\lambda_i$ is its $i$-th largest eigenvalue. Note that each row sum of $(1-\beta)\cdot\PtM+\beta\cdot\PaM$ is equal to $1$ and each entry of $(1-\beta)\cdot\PtM+\beta\cdot\PaM$ is non-negative. Then, by Lemma \ref{lem:stoch-eigen},  we have $\lambda_i\in [-1,1]$ for $1\le i\le n$. Let $f(\lambda_i)=\sum_{\ell=0}^{t}{\alpha(1-\alpha)^{\ell}\lambda_i^\ell}$. 
Lemma \ref{thrm:lambda} implies that any eigen-pair $\forall{i}\in [1,n], [f(\lambda_i),\mathbf{x}_i]$ of $(1-\beta)\cdot\PtM+\beta\cdot\PaM$ is an eigen-pair of $\SM$. By the sum of geometric sequence, we have $f(\lambda_i)=\alpha\cdot\frac{1-(1-\alpha)^{t+1}\lambda^{
t+1}_i}{1-(1-\alpha)\lambda_i}=\frac{\alpha}{1-(1-\alpha)\lambda_i}$,
which is is monotonously decreasing when $1\le i\le n$. Hence, for $1\le i\le n$, $f(\lambda_i)$ and $\mathbf{x}_i$ are the $i$-th largest eigenvalue and the $i$-th largest eigenvector of $\SM$. 
Recall that $\FM$ is the top-$k$ eigenvectors of $(1-\beta)\cdot\PtM+\beta\cdot\PaM$. Therefore, $\FM$ is the top-$k$ eigenvectors of $\SM$. The lemma is proved.
\done

\header
{\bf Proof of Lemma \ref{lem:svd}.}
Let $\mathbf{Z}=\mathbf{V}^{\top}\XM^{\top}\mathbf{U}$. Since $\mathbf{U}$ and $\mathbf{V}$ are the left and right singular vectors, we have $\mathbf{U}\mathbf{U}^{\top}=\IM$ and $\mathbf{V}\mathbf{V}^{\top}=\IM$. Note that $\ZM\ZM^{\top}=\IM$, which implies that each  $\ZM[i,j]$ satisfies $-1\le \ZM[i,j]\le 1$. Also, $\boldsymbol{\Sigma}[i,i]$ is a singular value and thus $\boldsymbol{\Sigma}[i,i]>0$. Then,
\begin{align*}
\textrm{trace}((\YM\YM^{\top})^{-\frac{1}{2}}\YM\FM^{\top}\XM^{\top})&=\textrm{trace}(\mathbf{U}\boldsymbol{\Sigma}\mathbf{V}^{\top}\XM^{\top})=\textrm{trace}(\boldsymbol{\Sigma}\mathbf{V}^{\top}\XM^{\top}\mathbf{U})\\
&=\textstyle\sum_{i=1}^{k}{\boldsymbol{\Sigma}[i,i]\cdot \mathbf{Z}[i,i]}\le \sum_{i=1}^{k}{\boldsymbol{\Sigma}[i,i]}.
\end{align*}
Therefore, $\textrm{trace}((\YM\YM^{\top})^{-\frac{1}{2}}\YM\FM^{\top}\XM^{\top})$ is maximized when $\ZM=\IM$, which implies that $\XM=\mathbf{U}\mathbf{V}^{\top}$. The lemma is proved.\done

\section{Comparison with Spectral Clustering}

It is straightforward to apply the classic {\em spectral clustering} \cite{von2007tutorial} (dubbed as \asc) on $\SM$ (see Eq. \eqref{eq:sexact}) to generate clusters. Here, we theoretically analyse its major difference from \gasc .

Given an attributed input graph $G$, the number $k$ of clusters, and the random walk factors $\alpha,\beta$ as inputs, \asc runs with three phases: (i) computing the attributed random walk probability matrix $\SM$; (ii) finding the top-$k$ eigenvectors $\FM$ of the attributed random walk probability matrix $\SM$; and (iii) generating an NCI $\YM$ via \kmeans with $\FM$. Specifically, given the top-$k$ eigenvectors $\FM\in \mathbb{R}^{k\times n}$, mathematically, \kmeans finds $k$ disjoint clusters $C_1,C_2,\cdots,C_k$ of $G$ such that
\begin{equation*}
\textstyle \min_{C_1,C_2,\cdots,C_k}{\sum_{i=1}^{k}{\sum_{v_j\in C_i}{\lnorm{\FM[:,v_j]-\frac{1}{|C_i|}\sum_{v_l\in C_i}{\FM[:,v_l]}}^2_2}}},
\end{equation*}
which can be rewritten as an NCI-based matrix form, \ie
\begin{align}
&\textstyle \min_{\YM\in \mathbb{1}^{k\times n}}{\lnorm{\FM^{\top}-\YM^{\top}\XM}^2_F}\label{eq:kmeans},\\
&\textstyle \textrm{where}\ \XM=(\YM\YM^{\top})^{-1}\YM\FM^{\top}\label{eq:kmx}.
\end{align}
\kmeans solves Eq. \eqref{eq:kmeans}  by using the expectation-maximization (EM) algorithm. Initially, it sets $\YM$ as a random incicator matrix (\ie $\mathbb{1}^{k\times n}$) and computes $\XM$ via Eq. \eqref{eq:kmx} accordingly. Then, with $\XM$ fixed, each entry $\YM[c_i,v_j]$ for cluster $C_i$ and node $v_j$ is updated as
\begin{equation*}
\textstyle \YM[c_i,v_j]=\begin{cases}
\textstyle 1, &\quad\text{$i=\argmin{1\le l\le k}{\lnorm{\FM[:,v_j]-\XM[l]}^2_2}$},\\
\textstyle 0, &\quad\text{else}.
\end{cases}
\end{equation*}
After that, with $\YM$ fixed, it recomputes $\XM$ by Eq. \eqref{eq:kmx} accordingly. \kmeans repeats the above optimization process in an alternative and iterative manner until convergence. Suppose that we find the optimal $\YM$ to Eq. \eqref{eq:kmeans}. Then we have $\FM=\FM\YM^{\top}(\YM\YM^{\top})^{-1}\YM$, which implies that \asc actually finds an NCI $\YM$ that minimizes the following objective function
\begin{align*}
&\textstyle \frac{2}{k}\cdot\textrm{trace}(\FM\YM^{\top}(\YM\YM^{\top})^{-1}\YM\cdot(\IM-\SM)\cdot(\FM\YM^{\top}(\YM\YM^{\top})^{-1}\YM)^{\top}),\\
&\textstyle\quad\quad\quad \textrm{where $\FM$ is the top-$k$ eigenvectors of $\SM$},
\end{align*}
which is totally different from our objective function in Eq. \eqref{eq:objt}. In a nutshell, the NCI $\YM$ returned by \asc fails to provide us an attributed conductance that approaches the optimal attributed conductance $\phi^{*}$ of $G$.



\balance
\bibliographystyle{abbrv}
\bibliography{main}

\end{document}